\documentstyle[pre,aps,eqsecnum,multicol,epsf,rotate]{revtex}

\begin{document}
\draft

\title{Field Theory of Branching and Annihilating Random Walks} 

\author{John L. Cardy ${}^*$ and Uwe C. T\"auber ${}^+$}

\address{University of Oxford, Department of Physics --- 
	Theoretical Physics, \\ 1 Keble Road, Oxford OX1 3NP, U.K. \\
	${}^*$ and All Souls College, Oxford \\ 
	${}^+$ and Linacre College, Oxford}

\date{\today, submitted to J. Stat. Phys., OUTP97 14S}
\tighten 

\maketitle

\begin{abstract}
We develop a systematic analytic approach to the problem of branching
and annihilating random walks, equivalent to the diffusion--limited
reaction processes $2 \, A \to \emptyset$ and $A \to (m+1) \, A$,
where $m \geq 1$.  
Starting from the master equation, a field--theoretic representation
of the problem is derived, and fluctuation effects are taken into 
account via diagrammatic and renormalization group methods.
For $d > 2$, the mean--field rate equation, which predicts an active
phase as soon as the branching process is switched on, applies
qualitatively for both even and odd $m$, but the behavior in lower
dimensions is shown to be quite different for these two cases.
For even $m$, and $d$ near $2$, the active phase still appears
immediately, but with non--trivial crossover exponents which we
compute in an expansion in $\epsilon = 2 - d$, and with logarithmic
corrections in $d = 2$. 
However, there exists a second critical dimension $d_c' \approx 4/3$
below which a non--trivial inactive phase emerges, with asymptotic
behavior characteristic of the pure annihilation process. 
This is confirmed by an exact calculation in $d = 1$. 
The subsequent transition to the active phase, which represents a new
non--trivial dynamic universality class, is then investigated within a 
truncated loop expansion, which appears to give a correct qualitative
picture.
The model with $m = 2$ is also generalized to $N$ species of
particles, which provides yet another universality class and which is
exactly solvable in the limit $N \to \infty$. 
For odd $m$, we show that the fluctuations of the annihilation process
are strong enough to create a non--trivial inactive phase for all
$d \leq 2$. 
In this case, the transition to the active phase is in the directed
percolation universality class.
Finally, we study the modification when the annihilation reaction is
$3 \, A \to \emptyset$. When $m = 0 \ ({\rm mod} \ 3)$ the system is
always in its active phase, but with logarithmic crossover corrections
for $d = 1$, while the other cases should exhibit an directed
percolation transition out of a fluctuation--driven inactive phase. 
\end{abstract}





\section{Introduction}
 \label{introd}

Reaction--diffusion processes, in addition to being of direct physical 
interest, present a relatively simple class of non--equilibrium
stochastic processes in which the role of fluctuations, so important
to the study of equilibrium statistical physics, may be investigated
\cite{review}. Although such processes have been studied for many
years through various kinds of mean--field or self--consistent
approximations, more recently there has been some progress in taking
into account the effects of fluctuations through the use either of
exact methods or of systematic renormalization group theories
\cite{jcardy}, backed up by extensive numerical simulations. 

It has become clear that, once the constraints of detailed balance are 
abandoned, these systems may exhibit a greater richness of phenomena
than their counterparts close to equilibrium, leading, for example, to
new types of critical behavior characterized by new universality
classes. It might appear that the freedom allowed by giving up
detailed balance would lead to such an unacceptably large number of
possibilities as to make the concept of universality of limited
usefulness, and indeed, from the point of view of the proliferation of
microscopic cellular automaton models this would seem to be true. 
However, as with equilibrium phenomena, the introduction of a
coarse--grained description through a field--theoretic formalism makes
it possible, at least in principle, to make a universal classification
feasible.

All such coarse--grained descriptions of stochastic processes, whether
or not they satisfy detailed balance, divide the dynamics into two
parts: a deterministic evolution corresponding to the mean--field
dynamics, and the noisy effect of fluctuations ignored in such a
description. When detailed balance is present, the form of the noise
is prescribed, so that the field--theoretic version of the
coarse--grained description is narrowly determined. In the absence of
detailed balance this is not so, yet it may be shown that the
particular form of the noise chosen is crucial to determining the
universal behavior near a non--equilibrium critical point. One way to
avoid this ambiguity, which works very well for reaction--diffusion
processes, is to develop a field--theoretic description of the problem
which is exact, being directly derived from the underlying master
equation, and to perform coarse--graining, in terms of a derivative
expansion, only at a later stage. This method, developed some time ago
by Doi and by Peliti \cite{doi,peliti} (see also Ref.~\cite{grasch}),
relies on the use of a Fock space formalism very similar to that
employed for quantum many--body systems. More recently,
field--theoretic renormalization group methods have been
systematically applied to the dynamic field theories describing some
simple processes to obtain what is, in principle, a complete
characterization of their universal properties
\cite{leecar,howcar,reydro}. 

It is the purpose of this paper to describe such a field--theoretic
investigation, largely but not entirely based on renormalization group
methods, into a class of reaction--diffusion processes which are
characterized by their description in terms of {\em branching and
annihilating random walks}. (A very brief account of this work has been
presented in Ref.~\cite{cartau}.) These processes were originally
studied in the 1980s by Grassberger et al. \cite{celaut} as exceptions
to the apparent rule that systems exhibiting non--equilibrium
transitions from a trivial absorbing state into a non--trivial noisy
`active' state should be in the same universality class as 
{\em directed percolation \em(DP). While this terminology originally
refers to a realization of this universality class describing
percolation processes which are asymmetric with respect to a privileged
space dimension \cite{dirper}, more commonly it now applies to a
dynamic universality class in which this dimension is interpreted as
time. In terms of reaction--diffusion processes, an example of such a
system would be one with a single species of particle $A$, undergoing
diffusive behavior, single--particle annihilation $A \to \emptyset$,
and branching $A \to 2 A$. There is always a trivial absorbing state,
with no particles and consequently no fluctuations. For sufficiently
low branching rate, this is the only stationary state, but for larger
values of this rate, another, non--trivial `active' stationary state
appears. The DP transition occurs at the point where this state first
appears. The robustness of the DP universality class has been tested
by measuring its critical exponents in numerical simulations on many
different types of system, and may be understood in terms of the
simplicity and irreducibility of the dynamic field theory which
describes it \cite{dpftrg}. Indeed, this `Reggeon' field theory was
first studied in a completely different context of particle
physics. Among the non--equilibrium phase transitions described by
this universality class are those occurring in the contact process 
\cite{contct}, the dimer poisoning problem in the
Ziff--Gulari--Barshad model \cite{zgbmod}, and in certain 
autocatalytic reaction models \cite{autcat}.

The model studied by Grassberger et al. \cite{celaut} was a rather
complicated probabilistic cellular automaton in one space and one time 
dimension, with antiferromagnetic interactions in which the
`particles' actually appear as kinks or domain walls. It was observed
that, while this model appears to exhibit a transition from an
inactive, absorbing state to a non--trivial active state, the various
critical exponents are far from those of directed percolation. The
authors of Ref.~\cite{celaut} already realized that this may be a
result of the fact that these kinks are conserved modulo 2 in the
basic dynamics, but a subsequent field--theoretic analysis of this
problem appeared to shed no light on this discrepancy: rather it
suggested that the system should, once again, be in the DP
universality class, indicating that there was apparently no way of
taking into account the constraint of local `parity' conservation in a
field--theoretic treatment \cite{grassb}. One of the main points of
our analysis is to show where this argument went wrong, and why such
systems, at least in low dimensions, should be in a new universality
class. 

Subsequent work showed that similar effects occur in other systems with
such a conservation law, and suggested that, at least in one dimension,
they all fall into this new universality class, characterized by 
{\em branching and annihilating walks with an even number of offspring}
\cite{taktre,tkinui,evmjen,redner}. These are reaction--diffusion
systems with the underlying reaction processes $2 A \to \emptyset$ and
$A \to (m+1) A$, with $m$ even. In one dimension, when $m = 2$,
another realization of this universality class occurs in certain
classes of dynamic Ising models which violate detailed balance
\cite{celaut,grassb,kising}, where once again the `particles'
correspond to domain walls. In this case, the transition from the
inactive phase (no domain walls in the asymptotic stationary state) to
the active phase corresponds to the disordering transition of the
Ising spins \cite{norsim,drorac}. In addition, other realizations of
this universality class have been found and also been investigated
numerically \cite{basbro,hayehi}.

On the other hand, when $m$ is odd and there is no conservation law, one
expects to recover the DP exponents, and this appears to be the case, at
least in one dimension \cite{celaut,taktre,mondim,odmjen}. In higher
dimensions, the numerical situation is less clear. Takayasu and
Tretyakov, in rather short simulations, found no evidence for a
non--trivial transition for $d \geq 2$, for either $m$ even or odd
\cite{taktre}. This is certainly consistent with the picture obtained
from mean--field theory, as we shall see. But the situation in two
dimensions is, we believe, more subtle. It is just this dimension at
which the fluctuation effects become important in the pure
annihilation process $2 A \to \emptyset$, giving logarithmic
corrections to scaling, and they cannot therefore be ignored there. In
fact we find in our analysis that these fluctuation effects are
sufficient to drive the existence of an inactive phase for small but
finite values of the branching rate, at least for odd $m$, and hence
the existence of a subsequent transition in the active phase, which is
in the two--dimensional DP universality class. 

For $m$ even, on the other hand, the fluctuations are not effective
until a second, lower critical dimension which we estimate to be
$\approx \frac43$, and which is certainly above $d = 1$. The existence
of two critical dimensionalities make a systematic
$\epsilon$--expansion type of approach to this problem
infeasible. However, we have performed a truncated loop expansion
which has the correct fixed point structure to describe the new 
$d = 1$ universality class. One striking feature of this is the
existence of dangerous irrelevant variables which make the
identification of the various critical exponents in terms of RG
eigenvalues slightly delicate. In fact, we may only make this
identification strictly in the active phase, and our analysis leaves
open the possibility of violations of the scaling relations between
these exponents and those at the critical point and in the inactive
phase, which have been written down on the basis of simple scaling
arguments.  

Our aim, therefore, is to investigate the following competing reaction
processes of identical particles $A$, which perform a random walk with
diffusion constant $D$:
\begin{eqnarray}
	{\rm annihilation :} \quad &k \, A \to \emptyset \ , 
		&\quad {\rm rate} \ \lambda_k 	\ ,
 \label{parann} \\
	{\rm branching :} \quad &A \to (m + 1) \, A \ ,
		&\quad {\rm rate} \ \sigma_m	\ , 
 \label{parbra}
\end{eqnarray}
with $k, m$ integers, $k \geq 2$, and $m \geq 0$; `$\emptyset$'
denotes an inert state. Occasionally, we shall in addition allow for a
spontaneous creation of particles in pairs, 
\begin{equation}
	{\rm pair \ production :} \quad \emptyset \to 2 \, A \ ,
		\quad {\rm rate} \ \tau	\ .
 \label{parpro}
\end{equation}

The corresponding mean--field rate equation for the average density
$n(t)$, neglecting any local fluctuation effects, reads
\begin{equation}
	{d \, n(t) \over d \, t} = - k \, \lambda_k \, n(t)^k
		+ m \, \sigma_m \, n(t) + 2 \, \tau \ ,
 \label{mfrate}
\end{equation}
with an initial density $n(0) = n_0$. If we set the pair production
rate, which acts as an `external field', to zero, Eq.~(\ref{mfrate})
has the stationary solutions $n = 0$ (`inactive state') and 
\begin{equation}
	\tau = 0 \, : \quad n = n_s \equiv 
	\left( {m \, \sigma_m \over k \, \lambda_k} \right)^{1/(k-1)}
 \label{asyden}
\end{equation}
(`active state'). Provided the `mass' $m \sigma_m$ is positive
(and $n_0 > 0$), the solution of Eq.~(\ref{mfrate}) approaches the
asymptotic density $n_s$ {\em exponentially} for $t \to \infty$,
\begin{equation}
	\tau = 0 \, : \quad n(t) = {n_s \over 
	\Bigl( 1 + \left[ (n_s / n_0)^{k-1} - 1 \right] \, 
		e^{- (k-1) m \sigma_m t} \Bigr)^{1/(k-1)}} \ . 
 \label{mfbarw}
\end{equation}
On the other hand, if $\sigma_m = 0$, this exponential long--time
behavior is replaced by a {\em power--law} decay,
\begin{equation}
	\sigma_m = \tau = 0 \, : \quad n(t) = {n_0 \over \left[ 1 + 
	k (k-1) \, \lambda_k \, n_0^{k-1} \, t \right]^{1/(k-1)}} \ . 
 \label{mfpann}
\end{equation}
In this sense, $\sigma_m = \sigma_c = 0$ may be viewed as a critical
point, described by the following set of critical exponents
\begin{eqnarray}
	&\tau = 0 \, , \; \sigma_m = \sigma_c \, : \quad 
			&n(t) \propto t^{-\alpha} \ ,
 \label{alphdf} \\
	&\tau = 0 \, , \; \sigma_m \downarrow \sigma_c \, : \quad 
			&n_s \propto (\sigma_m - \sigma_c)^\beta \ ,
 \label{betadf}
\end{eqnarray}
characterizing the long--time decay at the critical point, and the
growth of the asymptotic density in the active phase, respectively;
according to Eqs.~(\ref{asyden}) and (\ref{mfpann}) their mean--field
values are $\alpha_0 = \beta_0 = 1 / (k-1)$. Taking into account the
diffusive propagation of local density fluctuations, two more
independent exponents may be introduced which describe the divergence
of the characteristic length and time scales as the critical point is
approached,
\begin{eqnarray}
	\tau = 0 \, , \; \sigma_m \to \sigma_c \, : \quad
		&&\xi \propto |\sigma_m - \sigma_c|^{-\nu} \ ,
 \label{nudefn} \\
	&&t_c \propto \xi^z \propto |\sigma_m - \sigma_c|^{-z \nu} \ ,
 \label{zdefin} \\
	\sigma_m = \sigma_c \, , \; \tau \to 0 \, : \quad
		&&\xi \propto |\tau|^{-\nu_\tau} \, , \quad
		t_c \propto |\tau|^{-z \nu_\tau} \ ,
 \label{nutdef}
\end{eqnarray}
with the mean--field values $\nu_0 = 1/2$, $\nu_{\tau 0} = 1 / (d+2)$,
and $z_0 = 2$.

These mean--field exponents of course originate in the scaling
dimensions of the parameters $D$, $\lambda_k$, $\sigma_m$, and $\tau$
of the model. Introducing a momentum scale $\kappa$, and measuring
times in terms of lengths squared,
\begin{equation}
	[ x ] = \kappa^{-1} \ , \quad [ t ] = \kappa^{-2} \ , \quad
	[ n ] = \kappa^d \ , 
 \label{scdim1}
\end{equation}
(as the particle density scales as an inverse volume in $d$
dimensions), Eq.~(\ref{mfrate}) implies the `naive' dimensions
\begin{eqnarray}
	&[ D ] = \kappa^0 \, , \quad &[ \tau ] = \kappa^{d + 2} \, ,
 \label{scdim2} \\
	&[ \sigma_m ] = \kappa^2 \, , \quad 
	&[ \lambda_k ] = \kappa^{2 - (k-1) d} \ . 
 \label{scdim3}
\end{eqnarray}
Therefore, both particle production processes appear to be 
{\em relevant} perturbations to the pure annihilation reaction, and
for $\sigma_m = \tau = 0$ the annihilation rate becomes dimensionless 
at the (upper) critical dimension $d_c = 2 / (k-1)$. Thus, one expects
the mean--field rate equation (\ref{mfrate}) to provide an at least
qualitatively correct description for $d > 2$ in the case $k = 2$, and
for $d > 1$ in the case $k = 3$. Because of the relevant perturbation
$\sigma_m$, the situation {\em at} the critical dimension already
requires a careful analysis, and certainly for $d < d_c$ fluctuation
effects need to be taken into account seriously. E.g., for the pure
annihilation process with $k = 2$ it is known {\em exactly} that
Eq.~(\ref{mfpann}) is replaced by the {\em slower} asymptotic decay
\begin{equation}
	k = 2 \, , \; d < 2 \ : \quad n(t) \propto t^{- d / 2} \ ,
 \label{flpann}
\end{equation}
due to strong particle {\em anti}correlations which cannot be
smoothened out by the diffusion in low dimensions
\cite{doi,peliti,leecar}. At $d = 2$ for $k = 2$ and $d = 1$ for 
$k = 3$, one finds logarithmic corrections to Eq.~(\ref{mfpann}). 
However, for $k \geq 4$ the above rate equation should be essentially
correct in all physical dimensions. In the bulk of this paper, we
shall therefore be concerned with the case $k = 2$, and only briefly
discuss the three--particle annihilation reaction ($k = 3$) combined
with branching in Sec.~\ref{3ABARW}.

The outline of the rest of this paper is as follows. After introducing 
the field--theoretic description of these processes, with an emphasis
on the subtleties which occur when correctly taking into account
processes in which a conservation law is satisfied, in
Sec.~\ref{2dBARW} we look close to the critical dimension $d = 2$ of
the pure annihilation reactions to discover whether the branching
processes are relevant (thus leading immediately to the active phase)
or otherwise. For $m$ even we find that they are in fact irrelevant
close to two dimensions, leading to a transition into the active phase
at zero branching rate, with, however, non--mean--field exponents
which may be evaluated within an $\epsilon$ expansion. This
calculation suggests, however, that this situation might be reversed
in sufficiently low dimensions, and in Sec.~\ref{1dBARW} we confirm
this by an {\em exact} calculation in one dimension, using both
field--theoretic and exact lattice methods. The subsequent
non--trivial transition into the active phase is described within the
truncated loop expansion in Sec.~\ref{evBARW}, and the various
subtleties connected with the unusual fixed point which emerges are
investigated. Comments are given on the feasibility of extending this
calculation to higher orders. In Sec.~\ref{NsBARW} a generalization to
$N$ species is introduced and solved, in the hope of elucidating the
case $N = 1$. However, it turns out that the behavior as soon as 
$N \geq 2$ is quite different, leading to yet another universality
class. In the next Section we turn to the case of $m$ odd, and use
diagram resummation and RG methods to argue that for $d \leq 2$ the
fluctuation effects drive the existence of a non--trivial inactive
phase, counter to the predictions of mean--field theory. After a brief
discussion in Sec.~\ref{3ABARW} of the corresponding set of problems
when branching is added to the three--particle annihilation process
$3A \to \emptyset$, we conclude with a summary of our results and
possible directions for future theoretical and numerical work in this
area. 


\section{Master equation and field theory}
 \label{masfth}

In order to include fluctuation effects systematically in a 
mathematical description of the diffusion--limited reactions 
(\ref{parann})--(\ref{parpro}), we first write down the 
corresponding master equation and then derive a field--theoretic
representation for it, following standard procedures
\cite{doi,peliti,grasch,celaut,leecar,jcardy}. The latter may then be
treated with standard methods like perturbation theory, and the
long--time and long--distance scaling behavior may be inferred via the
application of the renormalization group.

Master equations for the temporal evolution of the probability
distribution $P(\alpha;t)$ for a configuration $\alpha$ have the
general form 
\begin{equation}
	{d \, P(\alpha;t) \over d \, t} = 
		\sum_\beta R_{\beta \to \alpha} \, P(\beta;t) -
		\sum_\beta R_{\alpha \to \beta} \, P(\alpha;t) \ ,
 \label{genmas}
\end{equation}
where $R_{\alpha \to \beta}$ denotes the transition rate for the
process $\alpha \to \beta$. For random walkers on a lattice (with
lattice constant $b_0$) subject to the reactions
(\ref{parann})--(\ref{parpro}), any configuration is characterized by
the integer site occupation numbers $n_i$, $i = 1, \ldots, N$, with
$\sum_i n_i = N_A(t)$, and the master equation may be decomposed into
the balance equations for the diffusion, annihilation, branching, and
pair production processes at each site $i$,
\begin{equation}
	{d \, P(\{ n_i \};t) \over d \, t} = \sum_i \left[
	{\partial \, P(n_i;t) \over \partial \, t} \bigg\vert_D +
	{\partial \, P(n_i;t) \over \partial \, t}\bigg\vert_\lambda +
	{\partial \, P(n_i;t) \over \partial \, t} \bigg\vert_\sigma +
	{\partial \, P(n_i;t) \over \partial \, t} \bigg\vert_\tau 
							\right] \ ,
 \label{tmaseq}
\end{equation}
with
\begin{equation}
	{\partial \, P(n_i;t) \over \partial \, t} \bigg\vert_D = 
	{D \over b_0^2} \sum_{\{ e \}} 
	\Bigl[ (n_e + 1) \, P(\ldots,n_i-1,n_e+1,\ldots;t) - 
		n_i \, P(\ldots,n_i,n_e,\ldots;t) \Bigr] \ ,
 \label{dmaseq}
\end{equation}
where $\{ e \}$ denotes the set of nearest--neighbor sites adjacent to
$i$,
\begin{eqnarray}
	{\partial \, P(n_i;t) \over \partial \, t} \bigg\vert_\lambda
	&&= {\lambda_k \over b_0^{(k-1)d}} 
		\Bigl[ (n_i+k) (n_i+k-1) \ldots (n_i+1) \, 
			P(\ldots,n_i+k,\ldots;t) - \nonumber \\ 
	&&\qquad \qquad - n_i (n_i-1) \ldots (n_i-k+1) \, 
			P(\ldots,n_i,\ldots;t) \Bigr] \ ,
 \label{amaseq} \\
	{\partial \, P(n_i;t) \over \partial \, t} \bigg\vert_\sigma 
	&&= \sigma_m \Bigl[ (n_i-m) \, P(\ldots,n_i-m,\ldots;t) -
			n_i \, P(\ldots,n_i,\ldots;t) \Bigr] \ ,
 \label{bmaseq}	
\end{eqnarray}
and finally
\begin{equation}
	{\partial \, P(n_i;t) \over \partial \, t} \bigg\vert_\tau =
	\tau \Bigl[ P(\ldots,n_i+2,\ldots;t) - P(\ldots,n_i,\ldots;t)
							\Bigr] \ . 
 \label{pmaseq}
\end{equation}
The precise form of the initial state is not important as we shall
largely be considering stationary state properties. For convenience it
is simplest to choose
an uncorrelated Poisson distribution
\begin{equation}
	P(\{ n_i \};t) = e^{- N_A(0)} \, 
			\prod_i \, {{\bar n_0}^{n_i} \over n_i !} \ ,
	\quad {\bar n}_0 = {N_A(0) \over N} \ .
 \label{initst}
\end{equation}

As the configurations are given entirely in terms of the occupation
numbers $n_i$, this calls for a representation in terms of
second--quantized bosonic operators
\begin{equation}
	\bigl[ a_i , a_j^\dagger \bigr] = \delta_{ij} \ , \quad
	a_i \, | 0 \rangle = 0 \ ,
 \label{boseop}
\end{equation}
whose effect on the state
\begin{equation}
	| \{ n_i \} \rangle = \prod_i \, (a_i^\dagger)^{n_i} \, 
							| 0 \rangle 
 \label{occvec}
\end{equation}
is to raise or lower the site occupation number $n_i$ by one,
respectively:
\begin{eqnarray}
	a_i^\dagger | \dots n_i \ldots \rangle &&= 
				| \ldots n_i+1 \ldots \rangle \ ,
 \label{ladop1} \\
 	a_i | \dots n_i \ldots \rangle &&= 
			n_i \, | \ldots n_i-1 \ldots \rangle \ .
 \label{ladop2}
\end{eqnarray}
Then, upon defining the time--dependent state vector
\begin{equation}
	| \Phi(t) \rangle = 
	\sum_{\{ n_i \}} P(\{ n_i \};t) \; | \{ n_i \} \rangle \ ,
 \label{stavec}
\end{equation}
the master equation (\ref{tmaseq})--(\ref{pmaseq}) may be recast into
an `imaginary--time' Schr\"odinger equation
\begin{equation}
	{\partial \over \partial \, t} \, | \Phi(t) \rangle =
			- H \, | \Phi(t) \rangle \ ,
 \label{schreq}
\end{equation}
with the normal--ordered `hamiltonian'
\begin{eqnarray}
	H[\{ a_i^\dagger \},\{ a_i \}] = \sum_i H_i = \sum_i \Biggl(
	&&- {D \over b_0^2} \sum_{\{ e \}} a_i^\dagger (a_e - a_i) - 
	{\lambda_k \over b_0^{(k-1)d}} \, 
	\Bigl[ 1 - (a_i^\dagger)^k \Bigr] a_i^k + \nonumber \\ &&
	+ \sigma_m\, \Bigl[ 1- (a_i^\dagger)^m \Bigr] a_i^\dagger a_i
	+ \tau \, \Bigl[ 1 - (a_i^\dagger)^2 \Bigr] \Biggr) \ .
 \label{hamilt}
\end{eqnarray}
Using the initial state (\ref{initst}), the formal solution to
Eq.~(\ref{schreq}) reads
\begin{equation}
	| \Phi(t) \rangle = e^{- H t} \, | \Phi(0) \rangle =
	e^{- N_A(0)} \, e^{- H t} \, e^{{\bar n}_0 \sum_i a_i^\dagger}
						\, | 0 \rangle \ . 
 \label{forsol}
\end{equation}

Our aim is of course to compute time--dependent expectation values of
observables $A$, which may be defined in terms of the configuration
probability according to
\begin{equation}
	\langle A(t) \rangle = 
		\sum_{\{ n_i \}} A(\{ n_i \}) \, P(\{ n_i \};t) \ ;
 \label{expdef}					
\end{equation}
for this, we need the projection state
\begin{equation}
	\langle P | = \langle 0 | \, \prod_i \, e^{a_i} \ , \quad
				\langle P \, | \, 0 \rangle = 0 \ .
 \label{projst}
\end{equation}
For then, as a consequence of $\langle 0 | \, e^{a_i}$ being a left
eigenstate of $a_i^\dagger$ with eigenvalue $1$, one has 
$\langle P | \, Q(\{ a_i^\dagger \},\{ a_i \}) = \langle P | \, 
Q(\{ 1 \},\{ a_i \})$ for any normal--ordered polynomial $Q$ of the
ladder operators. Therefore, the expectation value (\ref{expdef}) may
be written as 
\begin{equation}
	\langle A(t) \rangle = 
	\langle P | \, A(\{ a_i \}) \, | \Phi(t) \rangle = e^{-N_A(0)} 
	\, \langle 0 | \, A(\{ a_i \})\, e^{\sum_i a_i} \, e^{-H t} \,
		e^{{\bar n}_0 \sum_i a_i^\dagger} \, | 0 \rangle \ . 
 \label{expval}
\end{equation}
Notice also that probability conservation requires
\begin{equation}
	0 \equiv 
	\langle 0 | \, e^{\sum_i a_i}\, H(\{ a_i^\dagger \},\{ a_i \})
	= \langle 0 | \, e^{\sum_i a_i} \, H(\{ 1 \},\{ a_i \}) \ ,
 \label{procon}
\end{equation}
and therefore the hamiltonian generally has to vanish when the
creation operators $a_i^\dagger$ are formally set to $1$. We remark
that in order to calculate `inclusive' probabilities, such as the
expectation value of the local density 
$n_i(t) = \langle a_i^\dagger(t) a_i(t) \rangle$, it is convenient to
commute the factor $e^{\sum_i a_i(t)}$ through the hamiltonian in
Eq.~(\ref{expval}), which is equivalent to a shift of all 
$a_i^\dagger \to 1 + a_i^\dagger$, because then one just has to
compute a vacuum expectation value of a normal--ordered product, for 
which Wick's theorem applies. Yet, this is not necessary for the
evaluation of `exclusive' quantities, as, e.g., the probability that
site $i$ is occupied by one particle, while all the other sites are
empty, 
$\langle \delta_{n_i,1} \, \prod_{j \not= i} \delta_{n_j,0} \rangle$.
Certainly this shift is not required for the renormalization of the
model, and may in fact be even dangerous (as we shall see below).

In order to arrive at a field--theoretic representation, we now view
the computation of (\ref{expval}) as a bosonic quantum many--particle 
problem with the hamiltonian (\ref{hamilt}), and again follow standard
procedures \cite{cstpin} to write the expectation value as a
coherent--state path integral
\begin{equation}
	\langle A(t_0) \rangle = 
	{\int \prod_i d {\hat \psi}_i d \psi_i \, A(\{ \psi_i \}) \, 
	e^{-S[{\hat \psi}_i,\psi_i;t_0]} \over \int \prod_i 
	d{\hat \psi}_i d \psi_i\, e^{-S[{\hat \psi}_i,\psi_i;t_0]}}\ ,
 \label{dscpin}
\end{equation}
with the effective action
\begin{equation}
	S[{\hat \psi}_i,\psi_i;t_0] = \sum_i \left( \int_0^{t_0} \! dt
	\left[ {\hat \psi}_i(t) \, {\partial \over \partial \, t}
	\psi_i(t) + H_i(\{ {\hat \psi}_i(t) \},\{ \psi_i(t) \})\right] 	 
	- \psi_i(t_0) - {\bar n}_0 \, {\hat \psi}_i(0) \right) \ .
 \label{dscact}
\end{equation}
Finally, we perform the formal continuum limit, using the (assumed)
inversion symmetry of the lattice,
\begin{eqnarray}
	&&\sum_i \to b_0^{-d} \int \! d^dx \, , \quad 
	\psi_i(t) \to b_0^d \, \psi({\bf x},t) \, , \quad
	{\hat \psi}_i(t) \to {\hat \psi}({\bf x},t) \ , \nonumber \\
	&&{\bar n}_0 \to b_0^d \, n_0 \, , \quad
	\sum_{\{ e \}} \, [\psi_e(t) - \psi_i(t)] \to 
		b_0^{d+2} \, \bbox{\nabla}^2 \psi({\bf x},t) \ ,
 \label{contlt}
\end{eqnarray}
which results in the action
\begin{eqnarray}
	S[{\hat \psi},\psi;t_0] = \int \! d^dx &&\Biggl[ 
		\int_0^{t_0} \! dt \Biggl( {\hat \psi}({\bf x},t) 
	\left[ {\partial \over \partial \, t} - D \bbox{\nabla}^2
				\right] \psi({\bf x},t) - \lambda_k \,
	\Bigl[ 1 - {\hat \psi}({\bf x},t)^k \Bigr] \psi({\bf x},t)^k +
	\nonumber \\ &&\qquad \qquad 
	+ \sigma_m \, \Bigl[ 1 - {\hat \psi}({\bf x},t)^m \Bigr] 
	{\hat \psi}({\bf x},t) \psi({\bf x},t) + \tau \, \Bigl[ 1 - 
	{\hat \psi}({\bf x},t)^2 \Bigr] \Biggr) - \nonumber \\ &&\quad
	- \psi({\bf x},t_0) - n_0\, {\hat \psi}({\bf x},0) \Biggr] \ . 
 \label{action}
\end{eqnarray}
According to Eq.~(\ref{dscpin}), $S$ has to be dimensionless, and
using Eq.~(\ref{scdim1}) we find the scaling dimensions of the fields
to be
\begin{equation}
	[ {\hat \psi}({\bf x},t) ] = \kappa^0 \, , \quad
	[ \psi({\bf x},t) ] = \kappa^d \ ,
 \label{scdimf}
\end{equation}
and thus recover Eqs.~(\ref{scdim2}) and (\ref{scdim3}). As mentioned
in Sec.~\ref{introd}, fluctuation effects are therefore expected to be
important in physical dimensions in the cases $k = 2$ ($d \leq 2$) and
$k = 3$ ($d \leq 1$). 

It is important to realize that in the above derivation of
Eq.~(\ref{action}), the form of the nonlinear terms followed directly
from the master equation, and {\em no} further assumptions had to be
made about these fluctuation contributions. On the other hand, the
mean--field rate equation may be recovered from this action by
considering the `classical' equations of motion for the fields
${\hat \psi}$ and $\psi$, as given by the stationarity conditions
\begin{eqnarray}
	&&0 = {\delta S \over \delta \psi} = - \left( {\partial 
	\over \partial \, t} + D \bbox{\nabla}^2 \right) {\hat \psi} 
	- k \lambda_k \Bigl[ 1 - {\hat \psi}^k \Bigr] \psi^{k-1} 
	+ \sigma_m \Bigl[ 1 - {\hat \psi}^m \Bigr] {\hat \psi} 
	- \delta(t - t_0) \ ,
 \label{statc1} \\
	&&0 = {\delta S \over \delta {\hat \psi}} = 
	\left( {\partial \over \partial \, t} - D \bbox{\nabla}^2
	\right) \psi + k \lambda_k \, {\hat \psi}^{k-1} \psi^k
	+ \sigma_m \Bigl[ 1 - (m+1) {\hat \psi}^m \Bigr] \psi
	- 2 \tau \, {\hat \psi} - n_0 \, \delta(t) \ .
 \label{statc2}
\end{eqnarray}
Obviously, Eq.~(\ref{statc1}) has the stationary homogeneous solution
${\hat \psi} = 1$, and inserting this into (\ref{statc2}) yields
\begin{equation}
	{\partial \, \psi({\bf x},t) \over \partial \, t} = 
	D \, \bbox{\nabla}^2 \, \psi({\bf x},t) - k \, \lambda_k \, 
		\psi({\bf x},t)^k + m \, \sigma_m \, \psi({\bf x},t)
				+ 2 \, \tau + n_0 \, \delta(t) \ ,
 \label{cgrden}
\end{equation}
which, upon identifying $\psi({\bf x},t)$ with a coarse--grained local
density (which is only possible on this mean--field level), is the
natural generalization of Eq.~(\ref{mfrate}).

Notice that for {\em even} $k$ {\em and even} $m$ the hamiltonian part
(in brackets) of the action (\ref{action}), which does not explicitly
depend on the final or initial state, is invariant under simultaneous
`parity' transformations of the fields,
\begin{equation}
	{\hat \psi}({\bf x},t) \to - {\hat \psi}({\bf x},t) \ , 
	\quad \psi({\bf x},t) \to - \psi({\bf x},t) \ .
 \label{parsym}
\end{equation}
Physically, this corresponds to the fact that for even $k$ and $m$ the
particle number is locally conserved modulo 2 by both the annihilation
and branching processes. However, if one expands the action
(\ref{action}) about the stationary solution according to
\begin{equation}
	{\hat \psi}({\bf x},t) = 1 + {\tilde \psi}({\bf x},t) \ ,
 \label{shiftf}
\end{equation}
corresponding to commuting the factor $e^{\sum_i a_i}$ through the
hamiltonian in Eq.~(\ref{expval}), one arrives at the new action
\begin{eqnarray}
	S[{\tilde \psi},\psi;t_0] = \int \! \! d^dx &&\Biggl[ 
	\int_0^{t_0} \! \! dt \Biggl( {\tilde \psi}({\bf x},t) 
	\left[ {\partial \over \partial t} - D \bbox{\nabla}^2 - 
				m \sigma_m \right] \! \psi({\bf x},t) 
	+ \lambda_k \sum_{l = 1}^k {k \choose l}  
	{\tilde \psi}({\bf x},t)^l \, \psi({\bf x},t)^k - \nonumber \\
	&&\qquad \qquad -\sigma_m \sum_{l = 2}^{m+1} {m+1 \choose l}
		\, {\tilde \psi}({\bf x},t)^l\, \psi({\bf x},t) 
	- 2 \tau \, {\tilde \psi}({\bf x},t) - \tau \, 
	{\tilde \psi}({\bf x},t)^2 \Biggr) - \nonumber \\ && \quad 
	- \psi({\bf x},0) - n_0\, {\tilde \psi}({\bf x},0) \Biggr] \ , 
 \label{shfact}
\end{eqnarray}
in which the symmetry (\ref{parsym}) for even $k$ and $m$ is obscured. 
Even worse, if in accordance with usual naive power counting arguments
only the three--point vertices are kept and all the higher
nonlinearities are disregarded, this hidden symmetry is lost
completely. (Reasoning along these lines led a previous investigation 
to the erroneous conclusion that a field--theoretic treatment would
predict that the dynamic phase transition for BARW
with $k = 2$ should be in the directed percolation universality class 
{\em irrespective} of the parity of $m$ \cite{grassb}.) For a
consistent renormalization group analysis, it is of course imperative
to preserve all the symmetries of the problem. In the present case,
this may be done by observing that the RG equations themselves (as
opposed to the calculations of observables such as the density) should
be independent  of which basis is used, and it is therefore possible,
and, indeed, necessary, to perform the computations in the
representation of the model in which the symmetry for even $k$ and $m$
is manifest, namely the unshifted action (\ref{action}).


\section{BARW with two--particle annihilation near two dimensions}
 \label{2dBARW}

\subsection{RG eigenvalues at the annihilation fixed point and
	generation of new processes}
 \label{ncpgen}

We now begin the investigation of fluctuation effects for BARW with
two--particle annihilation ($k = 2$, $\lambda_2 \equiv \lambda$),
based on the action (\ref{action}). For $0 < t < t_0$, this field
theory is characterized by a propagator 
$\Theta(t) \, \exp\left[ -(D q^2 + \sigma) t \right]$ (in the time
domain) or $(-i \omega + D q^2 + \sigma_m)^{-1}$ (in the frequency
domain), and is graphically represented by a directed line, see
Fig.~\ref{mvert}(a), and by the vertices corresponding to the
annihilation [Fig.~\ref{mvert}(b,c)], branching [Fig.~\ref{mvert}(d)],
and pair production [Fig.~\ref{mvert}(e)] reactions, respectively. 
With these elements, a systematic perturbation expansion may be
constructed applying the usual techniques of quantum field theory
\cite{ftheor}. Taking into account fluctuation effects, we then define
the renormalization constants $Z_\lambda$, $Z_{\sigma_m}$, and
$Z_\tau$ according to
\begin{eqnarray}
	\ell &=& Z_\lambda \, \lambda \, C_d \kappa^{d-2} / D \ ,
 \label{renlam} \\
	s_m &=& Z_{\sigma_m} \, \sigma_m \, \kappa^{-2} / D \ ,
 \label{rensig} \\
	T &=& Z_\tau \, \tau \, \kappa^{-(d+2)} / D \ .
 \label{rentau}
\end{eqnarray}
The dimensionless renormalized couplings $\ell$ and $s_m$ introduced
here are to be determined from the vertex functions
$\Gamma_{{\hat \psi}{\hat \psi} \psi \psi}(-{\bf q}/2,-\omega/2;
-{\bf q}/2,-\omega/2;{\bf q}/2,\omega/2;{\bf q}/2,\omega/2)$ and
$\Gamma_{{\hat \psi}{\hat \psi}{\hat \psi} \psi}(-{\bf q}/2,-\omega/2;
-{\bf q}/2,-\omega/2;-{\bf q}/2,-\omega/2;3{\bf q}/2, 3\omega/2)$, 
respectively, at some specified symmetric normalization point setting
a momentum scale $\kappa$, e.g., $q^2/4 = \kappa^2$, $\omega = 0$, or 
${\bf q} = {\bf 0}$, $i \omega = 2 D \kappa^2$. In Eq.~(\ref{renlam}),
$C_d \equiv \Gamma(2-d/2) / 2^{d-1} \pi^{d/2}$ denotes a
$d$--dependent geometric factor ($C_1 = 1/2$, $C_2 = 1 / 2 \pi$). The
renormalized pair production rate $T$ may be obtained from the vertex
function $\Gamma_{{\hat \psi}{\hat \psi}}$ in the time domain, by
introducing a lower cutoff $t_i$ and taking the Laplace transform with
respect to $t_i$ at $i \omega = 2 D \kappa^2$. Similarly, the
renormalized annihilation and branching rates $\ell$ and $s_m$ may
also be computed from the vertex functions $\Gamma_{\psi \psi}$ and
$\Gamma_{{\hat \psi} \psi}$, respectively, in the time domain with an 
upper cutoff $t_f$. 

Near the upper critical dimension $d_c = 2$, we infer the scaling
behavior by studying first the ultraviolet (UV) divergences of the
perturbation theory, to be absorbed in the renormalization constants
$Z_\lambda$, $Z_{\sigma_m}$, and $Z_\tau$ defined in
Eqs.~(\ref{renlam})--(\ref{rentau}). From the solution of the RG
(Callan--Symanzik) equations for the correlation functions via the
method of characteristics, we derive running couplings given by the
differential RG flow equations
\begin{eqnarray}
	{d \, \ell(l) \over d \, l} = - \beta_\ell(l) \; , \quad
						&&\ell(0) = \ell \ ,
 \label{flolam} \\
	{d \, s_m(l) \over d \, l} = - \zeta_{s_m}(l) \, s_m(l) \; ,
					\quad &&s_m(0) = s_m \ ,
 \label{flosig} \\
	{d \, T(l) \over d \, l} = - \zeta_T(l) \, T(l) \; , \quad
						&&T(0) = T \ ,
 \label{flotau}
\end{eqnarray}
which yield the change of $\ell$, $s_m$, and $T$ under scale
transformations $\kappa \to \kappa \, e^{-l}$ (the asymptotic regime
being approached as $l \to \infty$). Here, the beta and zeta functions
follow from the $Z$ factors in Eqs.~(\ref{renlam})--(\ref{rentau})
according to
\begin{eqnarray}
	\beta_\ell(\ell) &=& 
	\kappa \, {\partial \over \partial \, \kappa} \, \lambda \ ,
 \label{betlam} \\
	\zeta_{s_m}(\ell) &=& \kappa \, {\partial \over 
		\partial \, \kappa} \, \ln {s_m \over \sigma_m} \ ,
 \label{zetsig} \\
	\zeta_T(\ell) &=& \kappa \, 
	{\partial \over \partial \, \kappa} \, \ln {T \over \tau} \ .
 \label{zettau}
\end{eqnarray}
A scale--invariant asymptotic regime is then described by a zero
$\ell^*$ of the beta function,
\begin{equation}
	\beta_\ell(\ell^*) = 0 \ ,
 \label{fpoint}
\end{equation}
and the corresponding power laws are given by the anomalous dimensions
(RG eigenvalues)
\begin{eqnarray}
	y_{\sigma_m} &=& - \zeta_{s_m}(\ell^*) \ ,
 \label{revsig} \\
	y_\tau &=& - \zeta_T(\ell^*)
 \label{revtau} 
\end{eqnarray}
at such a fixed point of the RG transformation, which is
infrared--stable if $\beta_\ell'(\ell^*) > 0$.

The obvious first question to be addressed is whether the particle
creation reactions with rates $\sigma_m$ and $\tau$ remain relevant
perturbations to the pure annihilation process, i.e., whether the
critical point remains at $\sigma_c = \tau_c = 0$, once fluctuation
effects are taken into account for $d \leq 2$ dimensions; and if so,
what the corresponding critical exponents are. To one--loop order, the
only UV--divergent Feynman diagrams at $d_c = 2$ which contribute to
the renormalization of the annihilation rate $\lambda$, the branching
rate $\sigma_m$, and the pair production rate $\tau$ are depicted in
Fig.~\ref{mreno}(a--c). These diagrams are identical with the
lowest--order contributions {\em at} the annihilation fixed point,
where $\sigma_m = \tau = 0$, provided the mass term $\propto \sigma_m$
in the propagator is set to zero. Upon using the dimensional
regularization and minimal subtraction scheme, one then finds the
following renormalization constants in $d = 2 - \epsilon$ dimensions,
\begin{eqnarray}
	Z_\lambda &=& 1 - {\lambda \over D} \, 
			{C_d \kappa^{-\epsilon} \over \epsilon} \ ,
 \label{2dzlam} \\
	Z_{\sigma_m} &=& 1 - {m (m+1) \over 2} \, {\lambda \over D} \, 
			{C_d \kappa^{-\epsilon} \over \epsilon} \ ,
 \label{2dzsig} \\
	Z_\tau &=& 1 - {\lambda \over D} \, 
			{C_d \kappa^{-\epsilon} \over \epsilon} \ .
 \label{2dztau}
\end{eqnarray}
The factor $m (m+1) / 2$ for $Z_{\sigma_m}$ simply originates in the
number of possibilities to select the external legs to the right of
the bubble in Fig.~\ref{mreno}(b).

From Eqs.~(\ref{renlam}), (\ref{betlam}), and (\ref{2dzlam}), we 
infer the beta function
\begin{equation}
	\beta_\ell(\ell) = \ell ( d - 2 + \ell ) \ .
 \label{annbet}
\end{equation}
In the pure annihilation model ($\sigma_m = \tau = 0$), this is
actually an {\em exact} result, which may be obtained either by
summing a geometric series of the bubble diagrams in
Fig.~\ref{mreno}(a), or by writing down the equivalent Bethe--Salpeter
equation for the annihilation vertex \cite{leecar}, leading to
\begin{equation}
	Z_\lambda^{-1} = 1 + {\lambda \over D} \, 
			{C_d \kappa^{d-2} \over 2 - d} \ ,
 \label{exzlam}
\end{equation}
and hence (\ref{annbet}). For $d > 2$ the
stable fixed point is just the `Gaussian' fixed point $\ell^* = 0$,
while for $d \leq 2$ the non--trivial annihilation fixed point
\begin{equation}
	\ell^* = \epsilon = 2 - d \; , \quad s_m^* = T^* = 0
 \label{pannfp}
\end{equation}
becomes stable and governs the asymptotic behavior. E.g., as there is
no renormalization of the propagator, and hence neither field nor
diffusion constant renormalization in the pure annihilation model, the
solution of the RG equation for the average density may be written as
\begin{equation}
	n(\kappa,\ell,t) = \kappa^d \, e^{- d l} \, 
	{\tilde n}\Bigl( t \, \kappa^2 \, e^{-2 l} \, \ell(l) \Bigr) \ .
 \label{panden}
\end{equation}
Near the fixed point (\ref{pannfp}), $\ell(l) \to \ell^*$, we then
match $e^{2l} = t \, \kappa^2$, and arrive at the already cited 
{\em exact} result (\ref{flpann}). Notice that in terms of the fixed
point (\ref{pannfp}) the renormalization constant can be written as
\begin{equation}
	Z_\lambda = 1 - \ell / \ell^* \ ,
 \label{afpzlm}
\end{equation}
and hence $\ell = \ell^*$ corresponds to an infinite bare annihilation
rate $\lambda$, c.f. Eq.~(\ref{renlam}) \cite{leecar}. Moreover,
Eqs.~(\ref{zetsig}), (\ref{zettau}), (\ref{2dzsig}), and
(\ref{2dztau}) yield the zeta functions
\begin{eqnarray}
	\zeta_{s_m}(\ell) &=& - 2 + {m (m+1) \over 2} \, \ell 
				+ {\cal O}(\ell^2) \ ,
 \label{2ztsig} \\
	\zeta_T(\ell) &=& - (d + 2) + \ell + {\cal O}(\ell^2) \ ,
 \label{2zttau}
\end{eqnarray}
and according to Eqs.~(\ref{revsig}) and (\ref{revtau}) the RG
eigenvalues for the branching and pair production processes at the
pure annihilation fixed point (\ref{pannfp}) become for $d \leq 2$
\begin{eqnarray}
	y_{\sigma_m} &=& 2 - {m (m+1) \over 2} \, \epsilon 
					+ {\cal O}(\epsilon^2) \ ,
 \label{2dysig} \\
	y_\tau &=& 4 - 2 \, \epsilon + {\cal O}(\epsilon^2) \ .
 \label{2dytau}
\end{eqnarray}

Thus, both these couplings remain indeed {\em relevant} near 
$d_c = 2$. However, it is important to note that in addition to the
original processes all the lower branching reactions with 
$m-2, m-4, \ldots$ offspring particles become generated via
fluctuations involving combinations of branching and annihilation
processes, see Fig.~\ref{mreno}(d). In a consistent RG treatment all
these additional reactions must be incorporated in the effective
action from the start. In the case of {\em odd} $m$, the most
important of these new processes is the {\em single--particle
annihilation} $A \to \emptyset$. Upon assigning to that the decay rate
$\mu$,
\begin{equation}
	{\rm decay :} \quad A \to \emptyset \, , \quad 
			{\rm rate}\ \mu\ ,
 \label{pardec}
\end{equation}
which has the scaling dimension
\begin{equation}
	[ \mu ] = \kappa^2
 \label{muscdm}
\end{equation}	
[compare Eq.~(\ref{scdim3})], the complete effective action reads
\begin{eqnarray}
	m \ {\rm odd} : \; S[{\hat \psi},\psi;t_0] = &&\int \! d^dx 
	\Biggl[ \int_0^{t_0} \! dt \Biggl( {\hat \psi}({\bf x},t) 
	\left[ {\partial \over \partial t} - D \bbox{\nabla}^2
				\right] \psi({\bf x},t) - \lambda \, 
	\Bigl[ 1 - {\hat \psi}({\bf x},t)^2 \Bigr] \psi({\bf x},t)^2 -
	\nonumber\\ &&-\mu \, \Bigl[ 1 - {\hat \psi}({\bf x},t) \Bigr]
	\psi({\bf x},t) + \sum_{l=1}^{(m+1)/2} \! \! \! \sigma_{2l-1}
		\, \Bigl[ 1 - {\hat \psi}({\bf x},t)^{2l-1} \Bigr] 
	{\hat \psi}({\bf x},t) \psi({\bf x},t) + \nonumber \\ 
	&&\qquad \qquad 
	+ \tau \, \Bigl[ 1 - {\hat \psi}({\bf x},t)^2 \Bigr] \Biggr) 
	- \psi({\bf x},t_0) - n_0\, {\hat \psi}({\bf x},0) \Biggr] \ . 
 \label{odmact}
\end{eqnarray}
The emergence of the spontaneous decay (\ref{pardec}) alters the
behavior of the system drastically, and in fact leads to a shift of
the critical point to $\sigma_c > 0$ for $d \leq 2$, with the ensuing
dynamic phase transition being in the directed--percolation
universality class, and $n(t)$ approaching its asymptotic value
exponentially in both the active and inactive (absorbing) phases. We
shall defer the discussion of the action (\ref{odmact}) and the
derivation of these results to Sec.~\ref{odBARW}.

For {\em even} $m$, on the other hand, just the lower even--offspring
branching processes are generated, and the effective action becomes
\begin{eqnarray}
	m \ {\rm even} : \; S[{\hat \psi},\psi;t_0] = \! \int \! d^dx
	&&\Biggl[ \int_0^{t_0} \! dt \Biggl( {\hat \psi}({\bf x},t) 
	\left[ {\partial \over \partial t} - D \bbox{\nabla}^2
				\right] \psi({\bf x},t) - \lambda \,
	\Bigl[ 1 - {\hat \psi}({\bf x},t)^2 \Bigr] \psi({\bf x},t)^2 +
	\nonumber \\ &&\qquad + \sum_{l=1}^{m/2} \sigma_{2l}
		\, \Bigl[ 1 - {\hat \psi}({\bf x},t)^{2l} \Bigr] 
	{\hat \psi}({\bf x},t) \psi({\bf x},t) + \tau \, \Bigl[ 1 - 
	{\hat \psi}({\bf x},t)^2 \Bigr] \Biggr) - \nonumber \\ &&\quad
	- \psi({\bf x},t_0) - n_0\, {\hat \psi}({\bf x},0) \Biggr] \ .
 \label{evmact}
\end{eqnarray}
However, Eq.~(\ref{2dysig}) implies that the most relevant of these
branching reactions is actually the one with smallest $m$, i.e., 
$m = 2$, with rate $\sigma_2 \equiv \sigma$ and RG eigenvalue
\begin{equation}
	y_\sigma = 2 - 3 \epsilon + {\cal O}(\epsilon^2) \ .
 \label{2dysg2}
\end{equation}
The branching process with $m = 2$ will therefore describe the entire
universality class of BARW with even offspring number $m$, clearly
distinct from the universality class of BARW with odd $m$ as a
consequence of the underlying `parity' symmetry (\ref{parsym}),
i.e., the conservation of particle number modulo 2. In the following
chapter, we shall derive the critical exponents for the BARW with 
$m = 2$ near two dimensions, and furthermore discuss the logarithmic
corrections induced by the marginality of the annihilation vertex in
$d_c = 2$.

\subsection{The case m = 2: Critical exponents near d = 2 and
	logarithmic corrections} 
 \label{2dlogs}
 
We have just argued that the generic universality class of the dynamic
phase transition for BARW with even offspring will be characterized by
the critical exponents for $m = 2$. To one--loop order 
[see Figs.~\ref{mreno}(a--c)], i.e., to first order in 
$\epsilon = 2 - d$, we found the RG eigenvalues (\ref{2dysg2}) and
(\ref{2dytau}) at the annihilation fixed point (\ref{pannfp}). Hence
both $\sigma$ and $\tau$ are relevant near $d_c = 2$, and the critical
point remains at its mean--field location $\sigma_c = \tau_c = 0$. 
Upon utilizing the mean--field result (\ref{asyden}), we write the
general solution of the RG equation for the density as
\begin{equation}
	n(\kappa,\ell,s,T,t) = 
	\kappa^d \, e^{-d \, l} \, {s(l) \over \ell(l)} \;
	{\hat n}\Bigl( t \, \kappa^2 \, e^{-2 l} \, \ell(l) \, ,
		s(l) / \ell(l) \, , T(l) / \ell(l) \Bigr) \ ;
 \label{rgdens}
\end{equation}
the pure annihilation result (\ref{panden}) is then recovered as 
$s \to 0$ and $T \to 0$, if we demand that 
${\hat n}(x,y,0) \to y^{-1}$ as $y \to 0$ and 
${\tilde n}(x) = \lim_{y \to 0} [y \, {\hat n}(x,y,0)]$. 
Therefore at the critical point the density decays as 
$n(t) \propto t^{-d/2}$, i.e., using Eq.~(\ref{alphdf}),
\begin{equation}
	\alpha = d / 2 \ .
 \label{2dalph}
\end{equation}

For nonzero branching and pair production rate, but in the vicinity of
the annihilation fixed point $\ell(l) \to \ell^*$, Eq.~(\ref{rgdens})
reduces to
\begin{equation}
	n(\kappa,\ell,s,T,t) \propto s \, e^{(y_\sigma - d) \, l} \; 
	{\hat n}\Bigl( t \, \kappa^2 \, e^{-2 l} \, \ell^* , \,
		s \, e^{y_\sigma l} , \, T \, e^{y_\tau l} \Bigr) \ .
 \label{2drgdn}
\end{equation}
As lengths scale as $e^l$, the correlation length diverges according
to $\xi \propto s^{-1 / y_\sigma}$ as $s \to 0$ and 
$\xi \propto s^{-1 / y_\tau}$ as $T \to 0$; therefore we identify
[see Eqs.~(\ref{nudefn}), (\ref{nutdef})]
\begin{equation}
	\nu = 1 / y_\sigma \; , \quad \nu_\tau = 1 / y_\tau \ .
 \label{2dnuex}
\end{equation}
Finally, the matching condition $e^{2 l} = t \kappa^2$ leads to
\begin{equation}
	n(\kappa,\ell,s,T,t) \propto s \, t^{(y_\sigma - d)/2} \;
	{\hat n}\Bigl( \ell^* , \, s \, t^{y_\sigma/2} , \,
					T \, t^{y_\tau/2} \Bigr) \ ,
 \label{2drgnt}
\end{equation}
which, upon comparison with (\ref{2dnuex}) and ({\ref{zdefin}})
immediately implies 
\begin{equation}
	z = 2 \ .
 \label{2ddyex}
\end{equation}
Furthermore, for $T = 0$ and $t \to \infty$ the $t$ dependence has to
cancel, i.e., ${\hat n}(\ell^*,x,0) \propto x^{- 1 + d / y_\sigma}$ as
$x \to \infty$. Hence with the definition (\ref{betadf}) we conclude
that
\begin{equation}
	\beta = d / y_\sigma = d \, \nu = z \, \nu \, \alpha \ .
 \label{2dbeta}
\end{equation}
Notice that the relations (\ref{2dalph}), (\ref{2dnuex}),
(\ref{2ddyex}) and (\ref{2dbeta}) are {\em exact} provided that
$\sigma_c = \tau_c = 0$. Only the RG eigenvalues $y_\sigma$ and
$y_\tau$ need to be computed perturbatively, e.g., in an expansion
near $d_c = 2$.

Yet, assuming that the one--loop result (\ref{2dysg2}) may be
extrapolated with reasonable accuracy to some finite value of
$\epsilon = 2 - d$, one finds that the branching reaction actually
appears to become {\em irrelevant} for $\epsilon > 2/3$, i.e., below a
supposedly {\em new critical dimension} 
\begin{equation}
	d_c' = 4/3 \ .
 \label{1lcrtd}
\end{equation}
Notice also that the exponents $\nu$ and $\beta$ diverge as $d_c'$ is
approached from above. This would suggest the possibility of an entire
{\em inactive phase} to emerge for $d < d_c'$ and $\sigma < \sigma_c$,
i.e., the dynamic phase transition would be shifted to some positive
critical value of the branching rate. For $\sigma < \sigma_c$ the
effective branching rate would then scale to zero under
renormalization, and the power laws of the pure annihilation theory
would apply. In such a situation the above scaling relations and
values of the critical exponents would have to be modified
considerably.

In order to pursue this issue further, we now determine the RG
eigenvalue $y_\sigma$ and $y_\tau$ at the annihilation fixed point to
second order in $\epsilon$. For that purpose, we need the
UV--divergent two--loop diagrams for the annihilation and branching
vertices, as depicted in Fig.~\ref{2dren}(a,b). Taking the external
momenta or frequencies at a symmetry point and using either of them to
fix the normalization scale, one may readily compute the Z factors for
$\lambda$ and $\sigma$, as defined in Eqs.~(\ref{renlam}) and
(\ref{rensig}); using dimensional regularization and minimal
subtraction, one eventually finds 
\begin{eqnarray}
	Z_\lambda &=& 1 - {\lambda \over D} \, 
		{C_d \kappa^{-\epsilon} \over \epsilon}
	+ {\lambda^2 \over D^2} \, 
		{C_d^2 \kappa^{-2 \epsilon} \over \epsilon^2} \ ,
 \label{2d2lzl} \\
	Z_\sigma &=& 1 - 3 \, {\lambda \over D} \, 
		{C_d \kappa^{-\epsilon} \over \epsilon} 
	+ 6 \, {\lambda^2 \over D^2} \, 
		{C_d^2 \kappa^{-2 \epsilon} \over \epsilon^2} 
	+ {3 \over 2} \, \ln {4 \over 3} \; {\lambda^2 \over D^2} \,
		{C_d^2 \kappa^{-2 \epsilon} \over \epsilon} \ .
 \label{2d2lzs}
\end{eqnarray}
Furthermore, by introducing a lower cutoff $t_i$ for the diagrams
contributing to the renormalization of the pair production rate, see
Fig.~\ref{2dren}(c), one finds
\begin{equation}
	Z_\tau = 1 - {\lambda \over D} \, 
		{C_d \kappa^{-\epsilon} \over \epsilon}
	+ {\lambda^2 \over D^2} \, 
		{C_d^2 \kappa^{-2 \epsilon} \over \epsilon^2} \ .
 \label{2d2lzt}
\end{equation}

As to be expected, Eqs.~(\ref{betlam}) and (\ref{2d2lzl}) result in
the same beta function (\ref{annbet}) as to one--loop order, for the
diagrams in Fig.~\ref{2dren}(a) are precisely those of the pure
annihilation theory, for which (\ref{annbet}) holds to all orders in
$\ell$, and therefore the fixed point is again $\ell^* = \epsilon$.
With Eqs.~(\ref{2d2lzs}) and (\ref{2d2lzt}), the zeta functions for
the branching vertex (\ref{zetsig}) and for the pair production rate
(\ref{zettau}) become
\begin{eqnarray}
	\zeta_s(\ell) &=& - 2 + 3 \, \ell 
	- 3 \, \ln {4 \over 3} \; \ell^2 + {\cal O}(\ell^3) \ ,
 \label{2lztsg} \\
	\zeta_T(\ell) &=& - (d+2) + \ell + {\cal O}(\ell^3) \ ,
 \label{2lztau}
\end{eqnarray}
and thus
\begin{eqnarray}
	y_\sigma &=& 2 - 3 \, \epsilon + 
	3 \, \ln {4 \over 3} \; \epsilon^2 + {\cal O}(\epsilon^3) \ ,
 \label{2d2lys} \\
	y_\tau &=& 4 - 2 \, \epsilon + {\cal O}(\epsilon^3) \ ,
 \label{2d2lyt}
\end{eqnarray}
with the RG eigenvalue for the pair production rate at the
annihilation fixed point being identical with the one--loop result
(\ref{2dytau}).

Eq.~(\ref{2d2lys}) for $y_\sigma$ would indicate that the branching
process becomes irrelevant for $d < d_c''$ with
\begin{equation}
	d_c'' = 2 - {1 - \sqrt{1 - {8 \over 3} \, \ln {4 \over 3}} 
		\over 2 \, \ln {4 \over 3}} \approx 1.1 \ ,
 \label{2lcrtd}
\end{equation}
and thus imply that in the interesting physical dimension $d = 1$ a
stable inactive phase exists. Of course, an expansion near the upper
critical dimension $d_c = 2$ can by no means conclusively justify such
an assertion. But we shall see in the following Sec.~\ref{1dBARW} that
a direct calculation in one dimension yields that surprisingly the
one--loop result (\ref{2dysig}) becomes exact (with $\epsilon = 1$),
and indeed the branching processes with even $m$ are irrelevant in 
$d = 1$ near the annihilation fixed point.

Yet, clearly in two dimensions all the previous results apply, and due
to the marginality of the annihilation vertex induce logarithmic
corrections to the mean--field critical exponents. With
Eqs.~(\ref{flolam}) and (\ref{annbet}) we find the running
annihilation rate 
\begin{equation}
	d = 2 \, : \quad \ell(l) = {\ell(1) \over 1 + \ell(1) \, l}\ , 
 \label{2dflol}
\end{equation}
and upon inserting this into Eqs.~(\ref{flosig}), (\ref{flotau}) and
(\ref{2ztsig}), (\ref{2zttau}) we may solve for the flows of $s$ and
$T$ as well,
\begin{eqnarray}
	d = 2 \, : \quad s(l) &=& 
	{s \, e^{2 \, l} \over \left[ 1 + \ell(1) \, l \right]^3} \ ,
 \label{2dflos} \\
	T(l) &=& {T \, e^{4 \, l}  \over 1 + \ell(1) \, l} \ .
 \label{2dflot}
\end{eqnarray}
We now employ Eq.~(\ref{rgdens}) to infer the scaling behavior in
the asymptotic limit $l \to \infty$, where $\ell(l) \propto l^{-1}$, 
$s(l) \propto e^{2 l}\, l^{-3}$, and $T(l) \propto e^{4 l}\, l^{-1}$,
and thus
\begin{equation}
	n(\kappa,\ell,s,T,t) \to \kappa^2 \, s \, l^{-2} \; 
	{\hat n}\Bigl( t \, \kappa^2 \ e^{-2 l} \, l^{-1},
		\, s \, e^{2 l} \, l^{-2} , \, T \, e^{4 l} \Bigr) \ .
 \label{2drgds}
\end{equation}

Consequently, upon identifying $\xi$ with $e^l$ again, and by matching
the second and third argument of ${\hat n}$ to one, respectively, we
find 
\begin{eqnarray}
	&T = 0 \, : \quad &\xi \propto s^{-1/2}\, \ln (1 / s) \ ,
 \label{2dlxi1}	\\
	&s = 0 \, : \quad &\xi \propto T^{-1/4} \ ;
 \label{2dlxi2}
\end{eqnarray}
in addition, from the prefactor of ${\hat n}$ we infer, using 
$2 l \approx \ln (1/s)$,
\begin{equation}
	T = 0 \, : \quad n_s \propto s / [\ln (1 / s)]^2 \ .
 \label{2dlbet}
\end{equation}
Finally, at the critical point $s = T = 0$, Eqs.~(\ref{2drgds}) or
(\ref{panden}) become
\begin{equation}
	n(\kappa,\ell,t) = \kappa^2 \, e^{-2 l} \;
	{\tilde n}\Bigl( t \, \kappa^2 \, e^{-2 l} \, l^{-1}\Bigr) \ .
 \label{2dlden}
\end{equation}
Matching the argument of ${\tilde n}$ to a constant then leads to
\begin{equation}
	s = T = 0 \, : \quad n(t) \propto t^{-1} \, \ln t \ ,
 \label{2dlalp}
\end{equation}
which is consistent with the solution of the mean--field rate equation
${\dot n} = - \lambda(t) \, n^2$ with a time--dependent annihilation
rate $\lambda(t) \propto 1 / \ln t$, see Eq.~(\ref{2dflol})
\cite{leecar}. For large initial densities $n_0 \gg n_s$, this 
logarithmic correction to the mean--field power--law relaxation may be
observable even for nonzero branching rate, in an intermediate time
window $(2 \, \lambda \, n_0)^{-1} \ll t \ll (2 \, \sigma)^{-1}$. 
Similarly, the non--trivial logarithmic corrections corresponding to
Eq.~(\ref{2dflos}) should appear in the opposite limit $n_0 \ll n_s$
for times $(2 \, \sigma)^{-1} \ll t \ll (2 \, \lambda \, n_0)^{-1}$,
inducing a subexponential increase
\begin{equation}
	n(t) \propto \exp 
	\left[ {2 {\tilde \sigma} t \over (\ln t)^3} \right] \ .
 \label{2dsbex}
\end{equation}

At very long times, however, any positive branching rate will
eventually lead to a purely exponential approach to $n_s$. This
crossover to an asymptotically Gaussian theory may actually be
described by simply keeping the mass term $\sigma$ in the propagator
for the one--loop diagrams depicted in Fig.~\ref{mreno}(a-c), which
then results in the following beta and zeta functions,
\begin{eqnarray}
	\beta_\ell(\ell,s) &=& \ell 
	\left( d - 2 + {\ell \over (1 + s)^{2 - d/2}} \right) \ ,
 \label{gaucrl} \\
	\zeta_s(\ell,s) &=& -2 + {3\, \ell \over (1 + s)^{2 - d/2}}\ ,
 \label{gaucrs} \\
	\zeta_T(\ell,s) &=& -(d + 2) + {\ell \over (1 + s)^{2 - d/2}}
 \label{gaucrt}
\end{eqnarray}
(for a related discussion of the crossover from a critical point to a
Gaussian theory, see Ref.~\cite{tausch}). The effective coupling
emerging here is
\begin{equation}
	g \equiv {\ell \over (1 + s)^{2 - d/2}} \ ,
 \label{effcpl}
\end{equation}
and its flowing counterpart $g(l)$ will vanish as $l \to \infty$
because of the divergence of $s(l)$, as long as $s$ remains a relevant
coupling. In this situation, i.e., for $d > d_c'$, the Gaussian fixed
point $g^* = 0$ describing the {\em active phase} is approached, and
the above flow functions (\ref{gaucrl})--(\ref{gaucrt}) assume their
mean--field values, see Sec.~\ref{evBARW}.

In summary, we have found that near two dimensions the BARW with even
offspring, described by the universality class of $m = 2$, are
characterized by a critical point at vanishing branching and pair
production rate, with critical exponents $\alpha = d / 2$, $z = 2$,
and only two non--trivial exponents $\nu$ and $\nu_\tau$, which we
have computed to second and first order in $\epsilon = 2 - d$,
respectively. We have also calculated the leading terms for the
logarithmic corrections appearing directly in two dimensions, and
identified the active phase with a Gaussian fixed point in terms of
the effective coupling (\ref{effcpl}). However, there remains the
possibility that in one dimension the branching process may actually
be irrelevant near the annihilation fixed point, opening up an entire
{\em inactive phase} for $0 \leq \sigma < \sigma_c$.


\section{BARW with two--particle annihilation and even number of
	offspring in one dimension}
 \label{1dBARW}

\subsection{Computation of $y_{\sigma_m}$ directly in d = 1}

In Sec.~\ref{2dBARW} it was shown that the RG eigenvalue
$y_{\sigma_m}$ which characterizes the relevance of the branching rate
at $\sigma = 0$ may be computed within the $\epsilon = 2 - d$
expansion, and, although it is relevant in $d = 2$, $y_{\sigma_m}$
starts to decrease as $\epsilon$ increases, indicating that it may
eventually change sign. In this section we compute its value exactly
in $d = 1$, by various methods, and show that this is indeed the case. 

In order to do this, we need to compute some dimensionless physical
quantity, at some particular length or time scale characterized by the
normalization scale $\kappa$, as a function of the bare parameters
$\sigma_m$ and $\lambda$, and understand how these latter should change
as $\kappa$ is varied, keeping the physical quantities fixed. Since we
are only interested in the ${\cal O}(s_m)$ terms in the RG equations
(where $s_m$ denotes the dimensionless counterpart of $\sigma_m$, see
Sec.~\ref{2dBARW}), we need to compute these physical quantities only
to first order in $\sigma_m$, although the computation should be
carried out, in principle, to all orders in the annihilation rate
$\lambda$, since for $d = 1$ $\lambda$ is not small at the
annihilation fixed point. Rather than the renormalized vertex
functions considered in Sec.~\ref{2dBARW}, we shall consider a
physically more accessible quantity, namely the probability
$P_{m+1}(t)$ that, given there is just one particle at $t = 0$, there
are exactly $m + 1$ particles found at time $t$. For sufficiently
small $t$, we have $P_{m+1}(t) \sim \sigma_m \, t$. Thus, we may
define a renormalized version of $\sigma_m$ in terms of the derivative
$dP_{m+1}(t)/dt$, evaluated at some later time $t = 1 / D \kappa^2$. 
This is clearly quite different from the definition used in
Sec.~\ref{2dBARW}; however, for universal quantities such as
$y_{\sigma_m}$, it should yield identical results. In order to make
this renormalized branching rate dimensionless, we then define 
\begin{equation}
	{\tilde s}_m \equiv t \, P_{m+1}'(t) \Big|_{t=1/D\kappa^2} \ ,
\end{equation}
where the tilde emphasizes the difference from the definition used
in Sec.~\ref{2dBARW}. 

$P_{m+1}(t)$ is given by the sum of all diagrams in the unshifted
theory given by Eq.~(\ref{action}), which, in time--ordered
perturbation theory, have exactly one particle at $t = 0$ and $m + 1$
particles when cut at time $t$. To first order in $\sigma_m$, there is
only one branching vertex. Even so, the set of diagrams which may
contribute to this may be very complicated. An important set is
illustrated in Fig.~\ref{zgzag}. It is clear that, once the initial
branching has taken place, the remaining portion of any such diagram
corresponds to a solution of the quantum ($m+1$)--body problem for
non--relativistic bosons interacting with a short--range repulsive
potential of strength $\lambda$. For general values of $d$ and
$\lambda$ this is intractable. However, in one dimension, and in the
limit $\lambda \to \infty$, these bosons should behave like free
fermions and therefore the problem be solvable.

Fortunately, it is just this limit which is appropriate to discuss the
behavior at the annihilation fixed point $\ell = \ell^*$, for this
corresponds to a divergent {\em bare} coupling $\lambda$. For $d < 2$
there are no ultraviolet divergences in the continuum theory, so that,
on dimensional grounds, in one dimension $P_{m+1}(t)$ has the form
$(\sigma_m / \lambda^2) \, F(\lambda^2 t / D) + {\cal O}(\sigma_m^2)$,
where $F$ is some scaling function. Hence [c.f. Eqs.~(\ref{rensig}), 
(\ref{zetsig})] ${\tilde Z}_{\sigma_m} = F'(\lambda^2 / D^2 \kappa^2) 
+ {\cal O}(\tilde s_m)$, 
and therefore
\begin{equation}
	\zeta_{{\tilde s}_m} = -2 \left[ 1 + 
	{\lambda^2 \over D^2 \kappa^2}\, {F''(\lambda^2 / D^2\kappa^2)
		\over F'(\lambda^2 / D^2 \kappa^2)} \right] \ .
\end{equation}
In this expression $\lambda$ is supposed to be expressed in terms of
the renormalized coupling by
\begin{equation}
	{\lambda \over 2 D \kappa} = {\ell^* \ell \over \ell^* - \ell}
								\ ,
 \label{infbcp}
\end{equation}
c.f. Eqs.~(\ref{renlam}) and (\ref{afpzlm}). From these equations we
see that, as $\ell \to \ell^*$, $\lambda \to \infty$, and that this is
the same as the limit $t \to \infty$.

This limit is more easily understood in the model defined on a
lattice. For then two particles annihilate with probability one when
they land on the same site. In this case, the branching process does
not make sense unless the offspring are placed on different sites. 
Thus, we consider a slightly different version of the model in which
the dimensionful parameter $\lambda$ is replaced by the lattice
spacing $b_0$. Once the $m$ particles are placed on neighboring sites,
they execute independent random walks until two or more of them fall
on the same site and they annihilate. Let $S_{m+1}(t)$ be the
probability that all $m+1$ particles, initially placed on neighboring
sites, survive until time $t$. Then we have the simple relation that
\begin{equation}
	P_{m+1}(t) = \sigma_m \int_0^t S_{m+1}(t-t') dt' 
			+ {\cal O}(\sigma_m^2) \ ,
\end{equation}
so that $\tilde s_m = \sigma_m S_{m+1}(t = 1 / D \kappa^2) 
+ {\cal O}(\sigma_m^2)$. 

The asymptotic behavior of $S_{m+1}$ is simple to compute. This
problem has been studied in $d=1$ in several contexts before
\cite{MEFisher}. Denoting the coordinates of the particles by
$(x_1,x_2,\ldots,x_{m+1})$, we observe that this vector undergoes an 
isotropic random walk in the region $x_{j+1} -x_j > 0$ of an 
($m+1$)--dimensional space, with absorbing boundaries along the
hyperplanes $x_{j+1} = x_j$, beginning at the point $x_j^0 = j b_0$. 
$S_{m+1}(t)$ is just the survival probability of this walk. The
Green's function for this problem is 
${\cal A}_{m+1} \, G^{(m+1)}(x_1-x_1^0,\ldots,x_m-x_m^0;t)$, where
${\cal A}_{m+1}$ is the completely antisymmetrizing operator on the
$m+1$ coordinates $x_j$, which ensures that the boundary condition is
satisfied, and $G^{(m+1)}$ is the usual ($m+1$)--dimensional lattice
Green's function on ${\rm Z}^{m+1}$. In the long--time limit we may
use a continuum approximation to this, with care: The survival
probability is then given by integrating the full Green's function
over the region $x_{j+1} > x_j$. The result must be a function of only
the scaled variables $(x_i^0 - x_j^0) / (Dt)^{1/2}$, each of which is
${\cal O}(b_0 / (Dt)^{1/2}) \ll 1$. The antisymmetry requires that the
lowest term in an expansion in these variables must have the form 
\begin{equation}
 \label{4a}
	\prod_{i<j} \Bigl[ (x_i^0 - x_j^0) / (Dt)^{1/2} \Bigr]
		\sim [b_0/(Dt)^{1/2}]^{m(m+1)/2} \ .
\end{equation}
This gives the result for finite lattice spacing, in the limit
$\lambda \to \infty$. The corresponding result in the continuum limit
for finite $\lambda$ is obtained by the simple replacement
$b_0 \to (2/\lambda)$. The precise coefficient (which is not, in any
case, relevant to the present calculation) may be obtained by studying
the case $m=2$, which is solvable in both limits.

Putting these results together, we then find that, in $d=1$,
\begin{equation} 
 \label{4b}
	y_{\sigma_m} = 2 - m(m+1)/2 \ ,
\end{equation}
{\it exactly}.

The form of the result in Eq.~(\ref{4a}) reflects the fermionic nature
of the particles for infinite $\lambda$. In fact, Eq.~(\ref{4b}) may
be derived more simply, if more formally, using this. Once the
branching event has occurred, the particles propagate as an
($m+1$)--body fermionic state. If we express this in terms of
anticommuting annihilation and creation operators $c_i$ and
$c^{\dag}_i$ on each lattice site, the branching term in the
hamiltonian has the form
\begin{equation}
	H_b = \sigma_m \sum_i \left( \prod_{j=-m/2}^{m/2}
					c^{\dag}_{i+j} \right) c_i 
\end{equation}
[for $m$ even; in the case of $m$ odd, the product runs from
$-(m+1)/2$ to $(m+1)/2$, omitting $j = 0$.] The continuum limit of
this expression is found by expanding in powers of $jb_0$. This will
be different from the bosonic case because the anticommuting nature of
the $c^{\dag}$ allows each derivative to appear only once. Thus the
lowest order, most relevant, term has the form
\begin{equation}
	b_0^{1+2+\ldots+m} \, c^{\dag} \, (\partial c^{\dag}) \,
		(\partial^2c^{\dag}) \ldots (\partial^mc^{\dag}) c \ ,
\end{equation}
so that this term is multiplied by an effective expansion parameter
$b_0^{m(m+1)/2} \sigma_m$, and its dimension is modified in a way
corresponding precisely to the second term in Eq.~(\ref{4b}).

\subsection{Results from an exactly solvable case}

Finally, we show how this result may, in the case $m = 2$, be derived
from an exactly solvable lattice version of the problem. This is the
model discussed by Takayasu and Tretyakov \cite{taktre}, in which a
site is chosen at random, and, if there is a particle there, it is
either moved to the left or right (with probability $p$), or two
offspring are placed on the neighboring sites (with probability $1-p$).
If the particles are interpreted as Ising domain walls, this model is
a discrete time version of a kinetic Ising model solved by Droz,
R\'acz, and Schmidt \cite{kising}, with Glauber dynamics at
effectively zero temperature, and Kawasaki dynamics at effectively
infinite temperature. In fact this model is always in an inactive
phase (except for $p = 0$), and does not exhibit the non--trivial
transition to an active phase which is one of our main concerns in
this paper. As shown in Ref.~\cite{norsim}, it is necessary to include 
another parameter in the model, which enhances the branching rate
relative to the diffusion process, in order to find such a
transition. Nevertheless, for the purposes of extracting the universal
eigenvalue $y_\sigma$, at the pure annihilation fixed point, this
deficiency is unimportant.

Takayasu and Tretyakov \cite{taktre} consider the probability $Q_r(t)$
that a given interval of length $r$ contains an even number of
particles. Since they assume translationally invariant initial
conditions, it does not matter which interval. In the Ising spin
language, this is simply related to the correlation function 
$\langle s_i \, s_{i+r}\rangle$. They show that this satisfies the
linear system of equations 
\begin{eqnarray}
	Q_r(t+1) - Q_r(t) &=& (1/N) \, \Big[ (2-p) 
		(Q_{r+1} - 2 Q_r + Q_{r+1}) \Big] \qquad (r \geq 2) \\
	Q_1(t+1) - Q_1(t) &=& (1/N)\, \Big[ (2-p) Q_2 - 2 Q_1 \Big]\ ,
\end{eqnarray}
where $N$ is the number of sites. Although Takayasu and Tretyakov did 
not analyze this equation in detail, it is not difficult to take a
continuum limit and extract the scaling behavior of the parameter 
$1 - p$, which is the analog of the branching rate 
$\sigma \equiv \sigma_2$. Introduce $Q_0$ so that the second equation
above has the same form as the first, extended down to $r = 1$. This
means taking 
\begin{equation}
	2 (1-p) (Q_1 - Q_0) = p Q_0 \ .
\end{equation}
On rescaling $t \to {\tilde t} = t/N$ and $r \to {\tilde r} = b_0 r$,
the continuum limit $1 / N \to 0$, $b_0 \to 0$ now gives a simple
diffusion equation 
\begin{equation}
	{\partial Q \over \partial t} = (2-p) \, b_0^2 \,
			{\partial^2 Q \over \partial r^2} \ ,
\end{equation}
for all real $r > 0$, together with boundary condition
\begin{equation}
	2 (1-p) \, b_0 \, {\partial Q \over \partial r} = p \, Q
\end{equation}
at $r = 0$. This corresponds then to a heat diffusion problem on the
half--line $r > 0$, with a radiating boundary at $r = 0$. There is a
length scale $L_b = (1-p) b_0$ implicit in the boundary condition. 
For $r \gg L_b$, the solution at late times will approach that with
the Dirichlet condition $Q(r=0) = 0$. However this will be modified
for $r \ll L_b$. The fact that $1 - p \sim \sigma$ scales with length
in this way immediately implies that $y_\sigma = -1$, consistent with
Eq.~(\ref{4b}). In fact, for this continuum model, this is true to all
orders in $\sigma$. This is because the non--trivial fixed point,
determining the transition at $p = 0$, is infinitely far away in this
scheme. 

Within this model, it is possible to determine $P_3(t)$ and hence
check the calculation given earlier in this section. If there is
initially only one particle in the system, then $Q_r(t=0) = r/N$. 
If $p = 1$ (no branching), this is an exact steady state solution,
corresponding to a uniform heat current. Let us find the solution in
perturbation theory in $1 - p \sim \sigma$, by writing 
$Q_r(t) = (r/N) + (1-p) f(r,t) + {\cal O}((1-p)^2)$. Then $f$ obeys
the diffusion equation, vanishes at $t = 0$, and satisfies
$2 (1-p) / N + {\cal O}((1-p)^2) = (1-p) f$ at $r = 0$. It is thus
given in terms of the Green's function $G_D(t-t';r,r')$, satisfying
Dirichlet boundary conditions at $r = 0$, by
\begin{equation}
	f(r,t) = {1 \over N} \int_0^t \! G_D(t-t';r,0) \, dt' 
	\sim {1 \over N} \int_0^t \! {r \over (t-t')^{3/2}} \, 
					e^{-r^2 / 4(t-t')} \, dt' \ .
\end{equation}
To first order in $\sigma$, there are only one or three particles in
the system, so that the density is
$Q_1(t) = (P_1 + 3 P_3)/N = (1 + 2 P_3)/N$. Hence 
$P_3(t) = (N/2) (1-p) f(r=b_0,t) + {\cal O}((1-p)^2)$, and the
${\cal O}(1-p)$ contribution to $P'_3(t)$ therefore scales as
$t^{-3/2}$. Following through the RG procedure described earlier then
leads to $y_{\sigma_2} = -1$ as before. 

Within this model, it is possible to investigate the nature of the
long--time behavior to all orders in $\sigma_2 \sim 1-p$. If $p < 1$,
the boundary temperature $Q_0(t)$ will increase to a point where the
heat current crossing the boundary equals that coming from large $r$. 
Thus $Q_0(t) \sim 2 (1-p)/N$, so that the density of particles, 
$Q_1(t) \sim [1 + 2 (1-p)/p]/N$. There is therefore a finite average
number of particles in the whole system, which diverges as $p \to 0$. 
It is easy to check that this is the asymptotic solution for any
finite odd initial number of particles. On the other hand, if this
number is even, then initially $Q_r(0) = r (N-r)/N$. In this case, the
only steady state solution is $Q_r = 0$, and it is straightforward to
check that the particle density $Q_1(t) \sim t^{-1/2}$, as expected in
the inactive phase.


\section{Two--offspring BARW: Truncated loop expansion}
 \label{evBARW}

\subsection{RG flows to one--loop order}
 \label{1lopRG}

In Sec.~\ref{2dlogs} we saw that near $d_c = 2$ dimensions, the
branching rate $\sigma$ for BARW with two offspring particles remains
a relevant perturbation. Thus the critical point is at $\sigma_c = 0$,
as in mean--field theory, and for any positive branching rate there
exists only the active phase, which is governed by a Gaussian fixed
point. On the other hand, the direct analysis of the one--dimensional
model in Sec.~\ref{1dBARW} unambiguously established the existence of
an inactive phase described by the power laws of the pure annihilation
model. Hence, in the inactive phase, $\sigma$ must constitute an
irrelevant operator in the RG sense.  Clearly, this behavior cannot be
constructed within an expansion about the upper critical dimension
$d_c=2$. Yet, by employing a {\em truncated loop expansion at fixed
dimension} we can establish a unifying RG framework for the above
results. To one--loop order, we shall indeed find a new critical
dimension $d_c' = 4/3$, below which the inactive phase emerges.

The complete set of tree and one--loop diagrams for the two-- and
four--point vertex functions is displayed in Fig.~\ref{1loop}. As
opposed to the analysis in Sec.~\ref{2dBARW}, we now retain the 
{\em full dependence} of both the vertices {\em and} the propagators
on the branching rate $\sigma$. Upon applying an upper cutoff $t_f$ in
time for the processes depicted in Figs.~\ref{1loop}(a,b), taking the
Laplace transform with respect to $t_f$ at the normalization point
$i \omega = 2 D \kappa^2$, and employing our previous definitions
(\ref{renlam}) and (\ref{rensig}), one readily arrives at the
renormalization constants
\begin{eqnarray}
	Z_\lambda &=& 1 - {C_d \over 2-d} \, {\lambda / D \over 
		\left( \kappa^2 + \sigma / D \right)^{1 - d/2}} \ ,
 \label{1lpdzl} \\
	Z_\sigma &=& 1 - 3 \, {C_d \over 2-d} \, {\lambda / D \over 
		\left( \kappa^2 + \sigma / D \right)^{1 - d/2}} \ .
 \label{1lpdzs}
\end{eqnarray}
Consequently, with Eqs.~(\ref{betlam}), (\ref{zetsig}) and 
$\zeta_\ell = \beta_\ell / \ell$, we find the zeta functions
\begin{eqnarray}
	\zeta_\ell(\ell,s) &=& d-2 + {\ell \over (1 + s)^{2-d/2}} \ ,
 \label{1lpzel} \\
	\zeta_s(\ell,s) &=& -2 + {3 \, \ell \over (1 + s)^{2-d/2}} \ .
 \label{1lpzes}
\end{eqnarray}
For $s = 0$, of course, these reduce to Eqs.~(\ref{annbet}) and
(\ref{2ztsig}) (for $m=2$) as computed at the annihilation fixed
point (\ref{pannfp}). Within this truncated one--loop expansion, we
find for the RG eigenvalue of $\sigma$ at this fixed point
\begin{equation}
	y_\sigma = 3d - 4 \ ,
 \label{trqlev}
\end{equation}
which indicates that indeed for $d < d_c' = 4/3$ the branching
processes become irrelevant. Note that in one dimension, this happens
to coincide with the exact result (\ref{4b}) for $m=2$. For 
$d > d_c'$, however, the annihilation fixed point is unstable with
respect to branching, and $s(l)$ will diverge as $l \to \infty$. In
this case we expect the flow of $\ell$ and $s$ to be described by
their naive scaling dimensions. Indeed, with 
$\ell(l) \sim e^{-(d-2)l}$ and $s(l) \sim e^{2l}$, we have 
$\ell(l) / s(l)^{2-d/2} \sim e^{-2l} \to 0$, and the fluctuation
contributions to Eqs.~(\ref{1lpzel}), (\ref{1lpzes}) vanish. In
Sec.~\ref{2dlogs}, we had already anticipated these flow functions
describing the crossover to the asymptotically Gaussian model that
characterizes the active phase.

Before discussing the ensuing flow equations in more detail, we
compute the renormalization of the pair production rate as well. 
Notice that in this calculation at fixed dimension, we cannot just
restrict ourselves to those diagrams that become ultraviolet divergent
at $d_c = 2$. The three graphs in Fig.~\ref{1loop}(c), computed with a
lower cutoff $t_i$, lead to the $Z$ factor
\begin{equation}
	Z_\tau = 1 - {C_d \over 2-d} \, {\lambda / D \over 
		\left( \kappa^2 + \sigma / D \right)^{1 - d/2}} 
	- {3 C_d \over 2-d} \, {\sigma \lambda \over D^2 \kappa^2}
	\left[ {1 \over (\kappa^2 + \sigma / D)^{1 - d/2}} - 
		{1 \over (\sigma / D)^{1 - d/2}} \right] \ .
 \label{1lpdzt}
\end{equation}
For $\sigma = 0$, this obviously reduces to Eq.~(\ref{2dztau}); on the
other hand, as $\sigma \to \infty$, we may expand the term in square
brackets with the result
\begin{equation}
	\kappa^2 \ll \sigma / D \ : \quad
	Z_\tau \to 1 - {C_d \over 2-d} \, {\lambda / D \over 
		\left( \kappa^2 + \sigma / D \right)^{1 - d/2}} + 
	{3 C_d \over 2} \, {\lambda / D \over (\sigma / D)^{1-d/2}} +
	{\cal O}\left({\lambda/D \over (\sigma/D)^{2-d/2}} \right) \ .
 \label{1ldzta}
\end{equation}
With the definition (\ref{zettau}) this leads to
\begin{equation}
	\zeta_T(\ell,s) = - (d+2) + {\ell \over (1 + s)^{2-d/2}} \ , 
 \label{1lpzet}
\end{equation}
which is identical to Eq.~(\ref{gaucrt}). Note that
Eq.~(\ref{1lpzet}), which has been obtained to leading order in an
expansion for large $\sigma / D \kappa^2$, correctly incorporates both
the inactive ($s = 0$) and active ($s \to \infty$) phases, as do
Eqs.~(\ref{1lpzel}) and (\ref{1lpzes}).

It is instructive to realize that one arrives at the same results 
for $\zeta_\ell$ and $\zeta_s$ by considering the four--point vertex
functions 
$\Gamma_{{\hat \psi}{\hat \psi} \psi \psi}(-{\bf q}/2,-\omega/2; 
-{\bf q}/2,-\omega/2;{\bf q}/2,\omega/2;{\bf q}/2,\omega/2)$ and
$\Gamma_{{\hat \psi}{\hat \psi}{\hat \psi} \psi}(-{\bf q}/2,-\omega/2;
-{\bf q}/2,-\omega/2;-{\bf q}/2,-\omega/2;3{\bf q}/2, 3\omega/2)$, 
respectively, at either the normalization point $q^2/4 = \kappa^2$, 
$\omega = 0$, or at ${\bf q} = {\bf 0}$, $i \omega = 2 D \kappa^2$.
This comes about because the expressions from the triangular loops in
Figs.~\ref{1loop}(d,e) may be written in the form of the last term of
Eq.~(\ref{1lpdzt}), and hence do neither contribute at the
annihilation nor at the Gaussian active fixed point. Finally, the loop
in the fourth diagram in Fig.~\ref{1loop}(d) carries no external
momentum or frequency, and thus has no influence on the zeta function.
As in the one--loop $\epsilon$ expansion near $d_c = 2$, there is
neither field renormalization nor renormalization of the diffusion
constant at this level, because the loop diagram for 
$\Gamma_{{\hat \psi} \psi}({\bf q},\omega)$ (with no upper cutoff in
the time domain), see Fig.~\ref{1loop}(b), carries no wave vector or
frequency dependence.

Returning to Eqs.~(\ref{1lpzel}), (\ref{1lpzes}), and (\ref{1lpzet}),
it is clear that the relevant effective coupling is indeed 
$g = \ell / (1+s)^{2-d/2}$, as defined in Eq.~(\ref{effcpl}). In the
prospective inactive phase, with $s \to 0$, we have $g \to \ell$, the
coupling of the pure annihilation model. On the other hand, if $s$
diverges, we may construct the beta function for $g$ as follows:
\begin{equation}
	s \to \infty \ : \quad \beta_g(g) \to 
	g \Bigl[ \zeta_\ell - \left( 2 - d/2 \right) \zeta_s \Bigr]
	= g \left[ 2 - {10 - 3d \over 2} \, g \right] \ .
 \label{1lbetg}
\end{equation}
Thus, as explained above, despite both $\ell(l)$ and $s(l)$ diverging
according to their naive scaling dimensions, there is a stable
Gaussian fixed point $g^* = 0$ describing the active phase. Yet,
Eq.~(\ref{1lbetg}) yields the additional non--trivial fixed point
\begin{equation}
	g^* = {4 \over 10 - 3 d} \ .
 \label{1lpfpt}
\end{equation}
As
\begin{equation}
	{d \, \beta_g(g) \over d \, g} = 2 - (10 - 3 d) \, g
 \label{1lbgpr}
\end{equation}
assumes the value $-2$ at $g = g^*$, this new fixed point turns out to
be infrared--unstable. As the fixed--point values for the new
effective coupling are limited by those of the annihilation rate,
which in turn has as upper bound $\ell^* = 2-d$ (for according to
Eq.~(\ref{infbcp}) this already corresponds to an infinitely large
bare coupling), $g^*$ enters the physical regime for 
$d \leq d_c' = 4/3$, precisely when the annihilation fixed point and
the associated inactive phase become stable. The non--trivial unstable
fixed point then governs the dynamical phase transition between the
power--law inactive and the Gaussian active phases.

The flow diagram in $d=1$, as obtained by numerically solving the flow
equations, is shown in Fig.~\ref{fld=1}. The flows in the upper left
half describe the active phase with both $s(l) \to \infty$ and
$\ell(l) \to \infty$, but the effective coupling $g$ approaching a
Gaussian fixed point, $g(l) \to 0$; on the other hand, the
trajectories in the lower right part of the figure correspond to the 
inactive phase, with $s(l) \to 0$ and 
$g(l) \to \ell(l) \to \ell^* = 2-d$ (annihilation fixed point) as
$l \to \infty$. Notice the remarkable flows in the inactive phase,
with the RG trajectories curling about the annihilation fixed point.
The unstable fixed point with $\ell(l) \propto s(l)^{2-d/2}$ (for
large $\ell$ and $s$) appears as a separatrix separating the basins of
attraction of the fixed points associated with the inactive and active
phase, respectively. For any dimension $d < d_c'$, the flow diagram
would look qualitatively similar. At $d = d_c' = 4/3$ (to this order)
itself, there appears an attractive line of fixed points with 
$\zeta_\ell = \zeta_s = 0$, starting at the point $\ell = 2/3$,
$s = 0$. For $d > d_c'$, all the flow trajectories tend to the
Gaussian fixed point $g^* = 0$ (with both $\ell, s \to \infty$). 
 
The ensuing phase diagram as function of space dimension is displayed
in Fig.~\ref{coupd}. For $d > d_c'$ and any nonzero branching rate
there exists only the active phase. For $d > d_c = 2$, the transition
at $\sigma_c = 0$ is described by the mean--field exponents, see
Sec.~\ref{introd}, while for $d_c' \leq d \leq 2$ one finds the
non--trivial critical exponents (logarithmic corrections at $d_c = 2$)
evaluated in Sec.~\ref{2dlogs}. Only for $d < d_c'$, when 
$g^* < \ell^* = 2-d$ and the annihilation fixed point becomes stable
for sufficiently small values of $s$, the inactive phase which is
governed by the power laws of the pure annihilation model emerges. The
critical behavior at the dynamic phase transition with $\sigma_c > 0$
is characterized by exponents belonging to a novel universality
class. In the following subsection, we shall analyze the scaling
behavior at this transition, and compute the critical exponents to
one--loop order. Although the actual values we find for these
exponents turn out to be rather poor estimates, as a consequence of
the absence of any small expansion parameter, we believe that the
qualitative features of the phase diagram, and specifically the
mechanism for how the inactive phase becomes possible, are correctly
encoded in the truncated loop expansion.

\subsection{Scaling analysis and critical exponents}
 \label{scalng}

We now want to explore the critical behavior in the vicinity of the
transition described by the unstable fixed point $g = g^*$,
Eq.~(\ref{1lpfpt}). We thus introduce
\begin{equation}
	\varepsilon = {g^* - g \over g^*} \ ,
 \label{pareps}
\end{equation}
and note that because of 
$g^* d\varepsilon(l) / dl = \beta_g(g(l)) = - g^* \varepsilon(l) \,
d\beta_g(g) / dg \big\vert_{g=g^*}$, we identify the corresponding
zeta function as
\begin{equation}
	\zeta_\varepsilon = 
	{d \, \beta_g(g) \over d \, g} \bigg\vert_{g=g^*} \ .
 \label{zeteps}
\end{equation}
Collecting the results (\ref{1lpzel}), (\ref{1lpzes}), (\ref{1lpzet}),
and (\ref{1lbgpr}), we then find the following RG eigenvalues at the
unstable fixed point $g^*$,
\begin{eqnarray}
	y_\varepsilon = - \zeta_\varepsilon^* &=& 2 \ ,
 \label{ztvefp} \\
	y_\lambda = - \zeta_\ell^* 
		&=& {(4 - d) (4 - 3 d) \over 10 - 3 d} \ ,
 \label{ztelfp} \\
	y_\sigma = - \zeta_s^* &=& {2 (4 - 3 d) \over 10 - 3 d} \ ,
 \label{ztesfp} \\
	y_\tau = - \zeta_T^* &=& {16 + 4 d - 3 d^2 \over 10 - 3 d} \ .
 \label{ztetfp}
\end{eqnarray}
We remark that $y_\lambda$ and $y_\sigma$ are not independent, but
fixed by the condition that $\beta_g(g^*>0) = 0$, and hence
$y_\lambda = (2-d/2) y_\sigma$. Notice also that to one--loop order
there are no anomalous dimensions for the diffusion constant or the
fields themselves.

In identifying these RG eigenvalues with the critical exponents,
however, we have to be careful as now the critical point is shifted
away from $\sigma_c = 0$ to $g_c = g^*$, and we therefore cannot apply 
Eqs.~(\ref{2dnuex}) and (\ref{2dbeta}). In general, the critical
exponents will rather depend on $y_\varepsilon$, as well as on
$y_\lambda$ and $y_\sigma$. Only the exponent $\nu_\tau$ is not
affected by these modifications, and therefore  
\begin{equation}
	\nu_\tau = 1 / y_\tau
 \label{1lopnt} 
\end{equation}
remains valid. In order to compute the exponents at $\tau = 0$, we
write the solution of the RG equation for the density in the form
\begin{equation}
	n(\kappa,\ell,s,t) = \kappa^d \, e^{-d \, l} \, {\tilde n} 
	\left( \kappa,\ell(l),s(l),t(\kappa^2/D)e^{-2l} \right) \ .
 \label{rgeden}
\end{equation}
By itself, this is not sufficient to determine the exponents, since
$s(l)$ and $\ell(l)$ do not themselves flow to fixed points and
the dependence of $n$ on these, even at early times, is not known
{\em a priori}. In the \em active \em phase, however, we know that
they do flow to a region in which mean--field theory is valid. This 
predicts that the right hand side of (\ref{rgeden}) has the form
$s(l) / \ell(l)$ times a function of the combination
$t(\kappa^2 / D) e^{-2l} s(l)$. 
In the vicinity of $g^*$, then, (\ref{rgeden}) becomes
\begin{equation}
	n(\kappa,\ell,s,t) = \kappa^d \, 
	e^{- ( d + y_\lambda - y_\sigma) \, l} \, {s \over \ell} \,
	{\bar n}\left( \varepsilon \, e^{y_\varepsilon \, l} \ ,
	t \, \kappa^2 \, e^{- (2 - y_\sigma) \, l} \, s \right) \ .
 \label{crdrge}
\end{equation}
Upon choosing the matching condition 
$e^{-l} = |\varepsilon|^{1/y_\varepsilon}$, this implies that we may
identify the critical exponents, as defined in Sec.~\ref{introd}, as
\begin{eqnarray}
	\beta &=& (d + y_\lambda - y_\sigma) / y_\varepsilon \ ,
 \label{1lopbt} \\
	\nu &=& (2 - y_\sigma) / 2 \, y_\epsilon \ ,
 \label{1lopnu} \\
	z &=& 2 \ .
 \label{1loopz}
\end{eqnarray}
Notice that lengths now scale as $e^{(1 - y_\sigma / 2) l}$. The
mean--field value for the dynamic exponent $z$ naturally follows from
the absence of diffusion constant renormalization. 

We emphasize that as a consequence of the appearance of the dangerous
irrelevant variable $1 / s$, we cannot unambiguously extract the
exponents defining the power laws {\em at} the critical point itself,
such as $\alpha$. In addition, this also precludes us to provide a
sound foundation based on the renormalization group for scaling
relations like $\beta = z \nu \alpha$ \cite{grassb,evmjen,norsim} for
the dynamic phase transition for $d < d_c'$. Such scaling relations
would follow only if the mean--field dependence $s(l)/\ell(l)$ were
also to be valid as both $s(l)$ and $\ell(l)$ grow without bound {\em
along} the critical line $g = g^*$. If the fixed point occurred at
finite values of these quantities, it would be reasonable to assume
that the right hand side of Eq.~(\ref{rgeden}) does have this
mean--field form, as it is to be evaluated at some finite rescaled
time. It would be interesting to look carefully for possible
violations of these scaling laws within simulations. 

Upon inserting Eqs.~(\ref{ztvefp})--(\ref{ztetfp}), we find the
following results as function of dimension, 
\begin{equation}
	\beta = {4 \over 10 - 3 d} \, , \ \nu = {3 \over 10 - 3 d} \,
	, \ \nu_\tau = {10 - 3 d \over 16 + 4 d - 3 d^2} \, , \ z = 2
	\ . 
 \label{1lcrex} 
\end{equation}
At the borderline dimension $d_c' = 4/3$, the value for the critical
exponent $\nu_\tau = 3/8$ coincides with the one found in the $\epsilon$
expansion about $d_c = 2$ (see Sec.~\ref{2dlogs}). On the other hand,
upon approaching $d_c'$ from below, one has $\nu = 1/2$ and 
$\beta = 2/3$, which has to be contrasted with $\nu, \beta \to \infty$
as $d \downarrow d_c'$. In the more interesting physical dimension 
$d = 1$ we find to one--loop order
\begin{equation}
	d = 1 \ : \quad y_\lambda = 3/7 \, , \ y_\sigma = 2/7 \, , \
			y_\tau = 17/7 \ ,
 \label{rgevd1}
\end{equation}
and thus
\begin{equation}
	d = 1 \ : \quad \beta = 4/7 \, , \ \nu = 3/7 \, , \ 
			\nu_\tau = 7/17 \, , \ z = 2 \ .
 \label{crexd1}
\end{equation}
These numerical values, compared to the actual simulation results
\cite{grassb,taktre,norsim,mondim} are generally rather poor, with the
remarkable exception of $\nu_\tau$. Keeping in mind that the truncated
loop expansion, with no small parameter at hand, constitutes an
uncontrolled approximation scheme, the unsatisfying accuracy may not
be too surprising.

\subsection{Some remarks regarding higher orders in the truncated loop
		expansion}
 \label{2lopRG}

It is by no means clear that including higher loop orders in the above
analysis would yield considerably better results for the critical
exponents, given that there is no a priori small expansion parameter
present at all. Yet some general remarks about a possible extension,
say, to two--loop order, are in place here. 

To two--loop order, one would obviously expect both field
renormalization and diffusion constant renormalization to appear,
e.g., from the frequency and wave vector dependence of the loop in the
fourth diagram of Fig.~\ref{2dren}(b), or more generally, from the set
of graphs depicted in Fig.~\ref{zgzag}. Defining
\begin{eqnarray}
	\zeta_\psi = \kappa \, {\partial \over \partial \, \kappa} \, 
			\ln {\psi_R \over \psi} \ ,
 \label{zetpsi} \\
	\zeta_D = \kappa \, {\partial \over \partial \, \kappa} \, 
			\ln {D_R \over D} \ ,
 \label{zetdif}
\end{eqnarray}
this will imply two additional RG eigenvalues $y_\psi = -\zeta_\psi^*$
and $y_D = - \zeta_D^*$. As may be inferred from Eqs.~(\ref{rgeden})
and (\ref{crdrge}), their appearance modifies the previous scaling
relations (\ref{1lopbt})--(\ref{1loopz}) to 
\begin{eqnarray}
	\beta &=& (d + y_\psi + y_\lambda - y_\sigma)/y_\varepsilon\ , 
 \label{2lopbt} \\
	\nu &=& (2 + y_D - y_\sigma)/ 2 \, y_\epsilon \ ,
 \label{2lopnu} \\
	z &=& {2 \, (2 - y_\sigma) \over 2 + y_D - y_\sigma} \ ,
 \label{2loopz}
\end{eqnarray}
and would thus allows for a non--trivial value of the dynamic exponent
as well.  

It is, however, far from obvious that an extension of the above
one--loop analysis to higher orders in the perturbation expansion is
feasible. Certainly, at the annihilation fixed point describing the
inactive phase, a loop expansion is well--defined to any order. Also,
the Gaussian fixed point is trivially described by the naive scaling
dimensions of the model parameters. Yet the unstable fixed point
governing the dynamic phase transition requires a subtle balancing of
the divergences of the couplings $s(l)$ and $\ell(l)$, as encoded in
the appearance of the new effective coupling $g(l)$. To one--loop
order, the expansion parameter in the bare theory is 
$(\lambda / D) / (\kappa^2 + \sigma/D)^{1-d/2}$, and hence the
effective coupling in the flow function becomes 
$g = \ell / (1+s)^{2-d/2}$. The two--loop diagrams, then, are
proportional to $(\lambda / D)^2 / (\kappa^2 + \sigma/D)^{2-d}$, and
thus their contributions to the zeta functions proportional to
$\ell^2 / (1+s)^{3-d} = g^2 (1 + s)$ --- which diverges at the
transition. It appears, therefore, that in an expansion in $1/s$, the
leading terms would have to cancel in order to render the loop
expansion well--defined. However, although it may be shown for 
$\zeta_\ell$ that this is fact true for those two--loop diagrams that
in the frequency domain are product of the one--loop terms, this
appears not to be the case for the nested loop diagrams such as the
fourth graph in Fig.~\ref{2dren}(b), and perhaps the summation of an
entire series of such diagrams, see Fig.~\ref{zgzag}, might be
indispensable. Despite some effort, we therefore did not succeed in
extending our one--loop analysis further, or even to demonstrate that
there exists a meaningful truncated two--loop theory at all.  


\section{Generalization to N species of particles}
 \label{NsBARW}

We now generalize the BARW with $k = m = 2$ to $N$ species of
particles $\alpha = 1,\ldots,N$, with only identical particles
annihilating each other, and two different types of branching
processes, 
\begin{eqnarray}
	{\rm annihilation :} \quad &A^\alpha + A^\alpha \to \emptyset
		\ , &\quad {\rm rate} \ \lambda \ ,
 \label{nspann} \\
	{\rm branching :} 
		\quad &A^\alpha \to A^\alpha + A^\alpha + A^\alpha \ ,
		&\quad {\rm rate} \ \sigma \ ,
 \label{nspibr} \\
	&A^\alpha \to A^\alpha + A^\beta + A^\beta \quad
	(\beta \not= \alpha) \ , &\quad {\rm rate} \ \sigma'/(N-1) \ ,
 \label{nspdbr} \\
	{\rm pair \ production :} 
		\quad &\emptyset \to A^\alpha + A^\alpha \ ,
		&\quad {\rm rate} \ \tau \ .
 \label{nsppro}
\end{eqnarray}
The motivation behind such a generalization is that we found in
Sec.~\ref{evBARW} that in the interesting physical case of $N = 1$ in
$d = 1$ the loop expansion was uncontrolled and does not lead to
reliable values for the exponents, even though it correctly predicts
the existence of a transition. As in the case of $N$--component
magnets, then, one might hope that there exists an $N$--species
generalization of the model with $m = 2$ introduced in
Sec.~\ref{masfth}, which becomes exactly solvable in the limit 
$N \to \infty$, yet which still retains the essential physics.
Unfortunately this does not seem to be so. There appears to be no
simple way of generalizing the action (\ref{evmact}) (with $m = 2$)
so that it possesses an $O(N)$ symmetry as in the magnetic case. The
highest symmetry that can be embedded while at the same time retaining
the conservation modulo 2 of each species is $S_N$, the permutation
symmetry of the $N$ species. This is clearly exhibited by the allowed
reactions above. Note, however, that the permutation symmetry allows
different possible rates for the reactions in (\ref{nspibr}) and
(\ref{nspdbr}). The symmetries of this model are most clearly seen in
the form of hamiltonian density in the second-quantized formalism:
\begin{equation}
	{\cal H}_{\rm int} = - \lambda \sum_\alpha 
	\left( 1 - {\hat \psi}_\alpha^2 \right) \psi_\alpha^2 
	- \sigma \sum_\alpha \left( {\hat \psi}_\alpha^2 - 1 \right) 
	{\hat \psi}_\alpha \psi_\alpha - {\sigma' \over N - 1} 
	\sum_{\alpha\not=\beta} \left( {\hat \psi}_\alpha^2 -1 \right)
				{\hat \psi}_\beta \psi_\beta \ . 
\end{equation}
When $\sigma'=(N-1)\sigma$, the branching terms do in fact exhibit an
$O(N)$ symmetry which generalizes the ${\rm Z}_2$ symmetry of the
original model. However, the annihilation terms inevitably violate it,
and any attempt to rectify this falls foul of the requirement of
probability conservation. This has the consequence that the symmetry 
of the branching terms is not preserved under renormalization. 
Physically, this is because after the first process (\ref{nspibr}),
any pair of the three outgoing particles may annihilate, so the
chances of observing three particles after some amount of time are
reduced as compared with process (\ref{nspdbr}), where only one pair
may annihilate. 

The two--loop calculation of Sec.~\ref{2dlogs} is readily generalized
to $N$ particle `flavors' by modifying the appropriate combinatorial
factors in the evaluation of the Feynman diagrams depicted in
Fig.~\ref{2dren}. Near two dimensions, this yields the $Z$ factors
(\ref{2d2lzl}), (\ref{2d2lzs}) as before, while the graphs in
Fig.~\ref{2dren}(b) for the new branching process (\ref{nspdbr}) give
\begin{equation}
	Z_{\sigma'} = 1 - {\lambda \over D} \, 
		{C_d \kappa^{-\epsilon} \over \epsilon}
+ {\lambda^2 \over D^2} \, 
		{C_d^2 \kappa^{-2 \epsilon} \over \epsilon^2} \ .
 \label{nspzbr}
\end{equation}
Consequently, with the definitions used in Sec.~\ref{2dBARW}, the beta
function (\ref{annbet}) and fixed point (\ref{pannfp}), as well as
Eq.~(\ref{2lztsg}) remain unaltered, while the additional zeta
function reads
\begin{equation}
	\zeta_{s'}(\ell) = - 2 + \ell + {\cal O}(\ell^3) \ ,
 \label{nspzts}
\end{equation}
and therefore the RG eigenvalue of the branching rate $\sigma'$ at the
annihilation fixed point is
\begin{equation}
	y_{\sigma'} = 2 - \epsilon + {\cal O}(\epsilon^3) \ .
 \label{nspys'}
\end{equation}
Upon comparing this result with (\ref{2d2lys}), we see that the
branching with identical offspring (\ref{nspibr}) is {\em irrelevant}
as compared to the production of different particle `flavors'
(\ref{nspdbr}): for $l \to \infty$, one has 
\begin{equation}
	s(l) / s'(l) \sim \exp\{ (y_\sigma - y_{\sigma'})\, l\} =
	\exp\{ [-2 \epsilon + 3 \ln (4/3) \epsilon^2]\, l \} \to 0 \ .
 \label{irrflo}
\end{equation}
Therefore the process (\ref{nspdbr}) is going to dominate the
long--time behavior, and we may effectively set $\sigma = 0$, but
retain $\sigma' > 0$, for any $N > 1$. The remaining Feynman diagrams
are then identical to those that would be obtained in the limit 
$N \to \infty$, and one therefore expects that actually all reactions
with $N > 1$ are asymptotically described by the model with infinitely
many particle species and sole branching process (\ref{nspdbr}).

The above model in the limit $\sigma = 0$, or equivalently, for
$N \to \infty$, has a considerably simpler structure as compared to
the $N = 1$ case, because `nested' diagrams as, e.g., the last graph
in Fig.~\ref{2dren}(b) do not appear any more. Consequently, all
fluctuation contributions to the propagator may be absorbed into a
renormalization of the `mass' $\sigma$. Therefore one is left with the
task of summing an infinite series of branching `bubble' diagrams
(`cactus' graphs). The most convenient way of achieving this is to
write down self--consistent Bethe--Salpeter type equations for the
annihilation, branching, and pair production rates. Graphically, these
are represented in Fig.~\ref{nspec}(a--c), respectively. The
corresponding analytic expressions read 
\begin{eqnarray}
	\lambda_R &=& -\, {1 \over 2}\, \Gamma_{\psi \psi}({\bf 0},0)
	= \lambda \left[ 1 - {\lambda_R \over D}
		\int_p {1 \over p^2 + \sigma_R'/D} \right] \ ,
 \label{ninflr} \\
	\sigma_R' &=& \Gamma_{{\hat \psi} \psi}({\bf 0},0) 
	= \sigma' \left[ 1 - {\lambda_R \over D}
		\int_p {1 \over p^2 + \sigma_R'/D} \right] \ ,
 \label{ninfsr} \\
	\tau_R &=& -\, {1 \over 2}\, 
			\Gamma_{{\hat \psi}{\hat \psi}}({\bf 0},0)
	= \lambda \left[ 1 - {\lambda_R \over D}
		\int_p {1 \over p^2 + \sigma_R'/D} \right] \ ,
 \label{ninftr}
\end{eqnarray}
which are readily solved by
\begin{equation}
	\lambda_R / \lambda = \sigma_R' / \sigma' = \tau_R / \tau 
	= \left[ 1 + {\lambda \over D} 
		\int_p {1 \over p^2 + \sigma_R'/D} \right]^{-1} \ .
 \label{ninfzt}
\end{equation}
If we now evaluate the integral at the normalization point 
$\sigma_R' / D = \kappa^2$, we arrive at the {\em exact} result
\begin{equation}
	Z_\lambda^{-1} = Z_{\sigma'}^{-1} = Z_\tau^{-1} = 
	1 + {\lambda \over D} \, {C_d \kappa^{d-2} \over 2 - d} \ ,
 \label{ninfzz}
\end{equation}
which immediately implies that Eqs.~(\ref{annbet}), (\ref{2zttau}),
and (\ref{nspzts}) hold to all orders in $\ell$,
\begin{equation}
	\beta_\ell(\ell) = \ell ( d - 2 + \ell ) \, , \quad
	\zeta_{s'}(\ell) = - 2 + \ell \, , \quad
	\zeta_T(\ell) = - (d + 2) + \ell \ .
 \label{ninfbz}
\end{equation}
Thus in this model, $\sigma'_c = \tau_c \equiv 0$, and the only
non--trivial RG eigenvalues are
\begin{equation}
	y_{\sigma'} = d \, , \quad y_\tau = 2 \, d \ ,
 \label{ninfys}
\end{equation}
which determine the divergence of the correlation length as either
$\sigma' \to 0$ or $\tau \to 0$.

We have thereby demonstrated that the above $N$--species
generalization of the BARW with $m = 2$ leads to a {\em new
universality class}, which for all $N > 1$ is asymptotically
characterized by the model with $N \to \infty$. In this model, the
critical point remains at vanishing branching and pair production
rate, and for $d < 2$ its critical behavior is governed by the
exponents 
\begin{equation}
	\nu = 1 / d \, , \quad \nu_\tau = 1 / 2 d \, , \quad 
	z = 2 \, , \quad \alpha = d / 2 \, , \quad \beta = 1 \ .
 \label{ninfex}
\end{equation}
In $d_c = 2$ dimensions, one finds a logarithmic correction for the
density decay at the critical point, just as in Eq.~(\ref{2dlalp}). 

The $N \to \infty$ limit of the model defined by the processes 
(\ref{nspann})--(\ref{nsppro}) may in fact be solved directly, without
going through the renormalization group machinery. This is based on
the simple observation that in the reaction 
$A^\alpha \to A^\alpha + A^\beta + A^\beta$, the two products
$A^\beta$ are far more likely to annihilate against each other, since
they have the same flavor index, than against the remaining
$A^\alpha$, or indeed any other particle in the system, which is
unlikely to have a matching flavor index. In the limit $N \to \infty$,
this becomes overwhelmingly the case. Now the average particle density
can change by branching processes, which will always be proportional
to the existing local density, and annihilation processes. These
latter are either between particles which are siblings, that is they
are products of the \em same \em branching event in the past, and
which will not, in this limit, annihilate against any other particle;
or they are between particles which are from independent branching
events. In this case they are unlikely to have the same flavor index,
so that this term is suppressed by a factor $1/N$; however, it is
proportional to the square of the mean density, which is itself
$O(N)$, and therefore it also gains a factor of $N$ relative to the
linear terms. If the particles come from independent branching events,
we may also, to leading order in $1/N$, neglect the correlation
between these events.  
 
One may therefore write down an integral equation for the
time evolution of the mean density $n(t)$:
\begin{equation}
 \label{largeNint}
	{dn \over dt} = 2 \sigma' \, n(t) - 2 \sigma' \int_0^t \!
		L(t-t') \, n(t') \, dt' - \lambda \int_0^t \! 
		\left[ 1 - L(t-t') \right] n(t')^2 \, dt' \ ,
\end{equation}
where $L(t-t') dt$ is the probability that two particles, created at
time $t'$, annihilate in the interval $(t,t+dt)$. Note that
$L(t) = - S'_2(t)$, where, as in Sec.~\ref{1dBARW}, $S_2(t-t')$ is the
survival probability at time $t$ for two isolated particles produced
at time $t'$. The factor of $1-L(t-t')$ in the last term ensures that
the particles have not annihilated at some earlier time.

Eq.~(\ref{largeNint}) may be solved simply by Laplace transform. 
Defining $\tilde n(p) = \int_0^\infty n(t) e^{-pt} dt$, and similarly
for $L(t)$, one finds
\begin{equation}
 \label{largeNLT}
	p \tilde n(p) - n(0) = \left[ 1 - \tilde L(p) \right] \left[ 
	2 \sigma' \tilde n(p) - \lambda \widetilde{n^2}(p) \right] \ ,
\end{equation}
where $\tilde L(p)$ is given by a simple sum of bubble diagrams as
\begin{equation}
	\tilde L(p) = {2 \lambda I(p) \over 1 + 2 \lambda I(p)} \ ,
\end{equation}
and $I(p) = \int_k 1 / (p + 2 D k^2)$. Although (\ref{largeNint})
cannot be solved in closed form, the various critical exponents may
easily be deduced. If we begin from an initial state when $n(0)$ is
small, we may initially neglect the nonlinear term to find that,
whenever $\sigma' > 0$, $\tilde n(p)$ has a pole for positive real $p$ 
at the solution $p = p_0(\sigma')$ of
\begin{equation}
	p \left[ 1 + 2 \lambda I(p) \right] = 2 \sigma' \ .
\end{equation}
The density therefore increases exponentially at a rate $e^{p_0 t}$.
When $d > 2$, $I(p)$ is analytic at $p = 0$ and we see that 
$p_0 \propto \sigma'$. When $d < 2$, $p_0 \propto {\sigma'}^{z \nu}$,
where $z \nu = 2 / d$, and for $d = 2$, 
$p_0 \sim \sigma' / \ln(1/\sigma')$, all in accord with the
renormalization group analysis presented above. For later times, this
exponential growth is capped by the nonlinear term and the system
reaches a steady state. Since the factor $1 - \tilde L(p)$ is common to
both terms on the right hand side of (\ref{largeNLT}), the steady state
density behaves simply as $\sigma^\beta$, where $\beta = 1$ irrespective
of the value of $d$.


\section{BARW with odd number of offspring} 
 \label{odBARW}

\subsection{Mapping to Reggeon field theory for directed percolation}

In this section, we consider the case of $m$ odd, which turns out to
be quite different from the case of even $m$. For odd $m$, there is no 
`parity' symmetry and one might expect the transition from an
inactive state, {\it provided} it occurs at some non--trivial value of
the branching rate $\sigma_m$, to be in the directed percolation (DP)
universality class, as a consequence of the generated particle decay
processes $A \to \emptyset$. This in fact is what we find in low
dimensions, but there are subtle issues connected with the fact that
these decay processes are themselves fluctuation--induced, and thus
proportional to the branching rate $\sigma_m$. Therefore the very
{\it existence} of a non--trivial transition must depend on
fluctuation effects, which are expected to be important only when 
$d \leq 2$. As pointed out in Sec.~\ref{introd}, the mean--field
equation always predicts an active state for all $\sigma_m > 0$, and
is expected to be valid, at least in the pure annihilation problem,
for $d > 2$. Indeed we find that only for dimensions $d \leq 2$,
fluctuations are strong enough to produce a nonzero critical branching
rate  $\sigma_c > 0$, and consequently a DP phase transition
separating the active from the inactive phase. This is despite the
fact that the upper critical dimension for the DP transition itself is
$d_c = 4$. 

For these reasons we investigate this problem in the vicinity of two
dimensions. As was shown in Sec.~\ref{2dBARW}, from an initial model
with a given value of $m>1$, under renormalization the other processes
with numbers of offspring $m-2,m-4,\ldots,1,-1$ are generated, compare
Fig.~\ref{mreno}(d). Therefore we shall begin by considering the
simplest case $m = 1$. We start by writing the interaction terms
appearing under the integral in the effective action
Eq.~(\ref{odmact}): 
\begin{equation}
 \label{Hint1}
	{\cal H}_{\rm int} = - \lambda \, (1 - {\hat \psi}^2) \, \psi^2 
        	- \sigma_1 \, ({\hat \psi} - 1) \, {\hat \psi} \psi
                - \mu ( 1 - {\hat \psi}) \psi \ .
\end{equation}
The last term corresponds to the process $A \to \emptyset$, which,
although not present in the original model, is generated by the
combined processes $A \to 2 A$, $2 A \to \emptyset$. 

It is also important to record the form of the interaction in terms of
the shifted field ${\tilde \psi} = {\hat \psi} - 1$. This is
\begin{equation}
 \label{Hint2}
	{\cal H}_{\rm int} = (\mu - \sigma_1) \, {\tilde \psi} \psi
	        + 2 \lambda \, {\tilde \psi} \psi^2
                - \sigma_1 \, {\tilde \psi}^2 \psi 
		+ \lambda \, {\tilde \psi}^2 \psi^2 \ ,
\end{equation}
which, apart from the last term, has the form of the effective
interaction for the (Reggeon) field theory of directed percolation
\cite{dpftrg}. With the simple rescaling
\begin{equation}
 \label{DPresc}
	\phi = \sqrt{2 \lambda / \sigma_1} \ , \quad
	{\tilde \phi} = \sqrt{\sigma_1 / 2 \lambda}
\end{equation}
one arrives at
\begin{equation}
 \label{Hint3}
	{\cal H}_{\rm int} = (\mu - \sigma_1) \, {\tilde \phi} \phi
        + u \left( {\tilde \phi} \phi^2 - {\tilde \phi}^2 \phi \right)
		+ \lambda \, {\tilde \phi}^2 \phi^2 \ ,
\end{equation}
with the new effective coupling for the three--point vertices
\begin{equation}
 \label{DPefcp}
	u = \sqrt{2 \lambda \sigma_1} \ , \quad 
	[ u ] = \kappa^{2-d/2} \ .
\end{equation}
Hence $u$ becomes marginal in $d = 4$, the upper critical dimension
for DP, and indeed one would argue that the last term in
Eq.~(\ref{Hint3}) should then be irrelevant close to the DP transition
by power counting. Thus, {\it provided} that $\sigma_{1c} > 0$, we
have shown that the ensuing dynamic phase transition from the inactive
to the active state is characterized by the critical exponents of the
DP universality class. Notice that {\it both} the actice and the
inactive phase are characterized by exponential long--time behavior,
as opposed to the power--law inactive phase for BARW with even $m$. At
first sight counterintuitively, the branching process {\em
accelerates} the particle density decay in the inactive phase, due to
the fluctuation--induced generation of spontaneous particle decay
processes.

Note, however, that in the bare theory the `mass' term (the
coefficient of ${\tilde \psi} \psi$) is {\it negative}, since 
$\mu = 0$ in that case. This would seem to imply that the model is
immediately in its active phase as soon as $\sigma_1 > 0$, and hence
the critical point at $\sigma_{1c} = 0$, which would render the
rescaling (\ref{DPresc}) obsolete. However, this argument neglects
renormalization effects. It may well be that the renormalized quantity
$\mu_R - \sigma_{1R}$ is positive. As we shall argue, this is in fact
the case for $d \leq 2$ and sufficiently small $\sigma_1$.

\subsection{Summation of the leading singularities near two
	dimensions} 

We are free to study the renormalization effects in either version of
the theory Eq.~(\ref{Hint1}) or Eq.~(\ref{Hint2}). In fact, this is
much simpler in the former, because it will turn out that the
effective expansion parameter is $\lambda$, and also because the mass
term in that case is $\mu + \sigma_1$, which is always positive,
allowing for a perturbative expansion with a meaningful propagator. Even
so, it is impossible to compute the renormalization to all orders. 
Instead, close to two dimensions we adopt the strategy of
retaining only the most singular diagrams as $\epsilon = 2-d \to 0$, 
at each order in $\lambda$. It is not difficult to see that these are
given by the iterated bubble diagrams shown in Fig.~\ref{buble}(a,b). 
Notice that all the propagators are renormalized by a series of
processes as depicted in Fig.~\ref{buble}(c), but all these side
branches may be resummed with the effect of replacing the mass in the 
internal propagator by its renormalized value. In the rest of this
section, we set the diffusion constant $D=1$, since it is not
renormalized at the order we consider. The ensuing geometric
sums then give the following simple self--consistent equations, to
leading order in $\epsilon$, 
\begin{eqnarray}
	\mu_R &=& \mu + {\sigma_1 (\lambda / 2 \pi \epsilon)
		(\mu_R + \sigma_{1R})^{-\epsilon/2} \over
		1 + (\lambda / 2 \pi \epsilon) 
		(\mu_R + \sigma_{1R})^{-\epsilon/2}} \ , \\
	\sigma_{1R} &=& {\sigma_1 \over
		1 + (\lambda / 2 \pi \epsilon) 
		(\mu_R + \sigma_{1R})^{-\epsilon/2}} \ .
\end{eqnarray}
Note that $\mu_R + \sigma_{1R} = \mu + \sigma_1$ in this
approximation. With $\mu = 0$ originally, this amounts precisely to
the summation of branched diagrams [see Fig.~\ref{buble}(c)] mentioned
above. However, we are interested in the renormalized value of the DP
mass (c.f. (\ref{Hint3}) 
\begin{equation}
 \label{DeltaR}
	\Delta_R \equiv \mu_R - \sigma_{1R} = \sigma_1 \, 
	{(\lambda / 2 \pi \epsilon) \sigma_1^{- \epsilon/2} - 1 \over
	(\lambda / 2 \pi \epsilon) \sigma_1^{-\epsilon/2} + 1} \ ,
\end{equation}
where we have used the above result, with $\mu = 0$. From
Eq.~(\ref{DeltaR}) we see that, for sufficiently small $\sigma_1$, the
DP mass is in fact {\it positive}, indicating that the system is in
fact in the {\it inactive} phase. The transition to the active phase
does not happen until $\sigma_1 = \sigma_{1c}$, where, in this
approximation, 
\begin{equation}
 \label{sigmac}
	\sigma_{1c} = (\lambda / 2 \pi \epsilon)^{2/\epsilon} \ .
\end{equation}
Although we would not expect this result to be quantitatively correct 
for $\epsilon = 1$, the power of $\lambda$ is exact and is dictated by
dimensional analysis. Thus we expect a DP transition at a non--trivial
value of $\sigma_1$ for all $d < 2$. This transition is driven by the
fluctuations in the annihilation process. 

Since the annihilation rate is (marginally) irrelevant exactly in two
dimensions, it is an interesting question whether it is able to drive a
DP transition in this case. The single bubbles are then logarithmically
divergent, and Eq.~(\ref{DeltaR}) is replaced by
\begin{equation}
	\Delta_R = \sigma_1 \, 
	{(\lambda / 4 \pi) \ln (\Lambda^2 / \sigma) - 1 \over
	          (\lambda / 4 \pi) \ln (\Lambda^2 / \sigma) + 1} \ ,
\end{equation}
where $\Lambda$ is a wave number cutoff (of the order of the inverse
lattice spacing $b_0^{-1}$). We see that indeed there is a
non--trivial transition, at 
\begin{equation}
	\sigma_{1c} \sim \Lambda^2 \, e^{-4 \pi / \lambda} \ .
 \label{crisig}
\end{equation}

This is one of the important results of this paper. It shows how
essential it is to take into account fluctuation effects in studying
this and similar processes. It appears to be in accord with recent
detailed simulations \cite{odmjen}, even though initial work
\cite{taktre} seemed to indicate that $\sigma_c$ might be zero.
Finally, for $d>2$, the same approximation (which is, however, less
justified), gives
\begin{equation}
	\Delta_R \sim \sigma_1 {\lambda \, I(\Lambda,\sigma_1) - 1
			\over \lambda \, I(\Lambda,\sigma_1) + 1} \ ,
\end{equation}
where 
\begin{equation}
	I(\Lambda,\sigma_1) = \frac14 \int_{|k|<\Lambda} \!
	{d^d\!k \over (2\pi)^d} \, {1 \over k^2+\sigma_1} \sim 
	\Lambda^{d-2} \qquad {\rm for} \ \sigma_1 \ll \Lambda^2 \ .
\end{equation}
In this case, as long as $\lambda \Lambda^{d-2} \ll 1$, the numerator
will be negative even at $\sigma_1 = 0$, and the system will be in the
active phase for all $\sigma_1 > 0$. There does appear to be the
possibility of a non--trivial transition at larger values of the
annihilation rate, when $\lambda \Lambda^{d-2} \gg 1$, but it should
be noted that our approximation of summing the leading bubble diagrams
breaks down long before then, and that it may be argued that 
{\it infinite} annihilation rate on the lattice already corresponds to
$\lambda \Lambda^{d-2} \sim 1$ in the continuum theory. There is no
evidence for a non--trivial transition in $d = 3$ from the simulations
\cite{taktre}.

\subsection{RG approach}

The same problem, with $m = 1$, may be attacked using the RG loop
expansion used in Sec.~\ref{2dBARW}. Once again, we work in the
unshifted theory given by Eq.~(\ref{Hint1}). The one--loop diagrams
contributing to the renormalization of $\lambda$, $\sigma_1$, and
$\mu$ are shown in Fig.~\ref{mreno}(a,b,d), and in fact only the 
combinatorial factors differ from the case $m = 2$ studied
above. Defining dimensionless renormalized couplings as in 
Eqs.~(\ref{renlam}), (\ref{rensig}), and in addition
\begin{equation}
	M = Z_\mu \mu \kappa^{-2} / D \ ,
 \label{renomu}
\end{equation}
we find the flow equations
\begin{eqnarray}
 \label{RG1}
	{d \ell(l) \over dl} &=& - \beta_\ell = \ell \left( \epsilon -
			{\ell \over (1+M+s_1)^{2-d/2}} \right) \ ,\\ 
	{d s_1(l) \over dl} &=& s_1 \left( 2 - 
			{\ell \over (1+M+s_1)^{2-d/2}} \right) \ , \\
	{d M(l) \over dl} &=& 2 \, M + 
			{\ell s_1 \over (1+M+s_1)^{2-d/2}} \ .
\end{eqnarray}
Note that the effective dimensionless mass appearing in the
denominators is now $M + s_1$. 

Adding the last two equations, we see that 
$M(l) + s_1(l) = \sigma_1(0) \, e^{2l}$ (equivalent to our earlier
result that the dimensionful mass does not renormalize). Although the
remaining equations cannot be solved by quadrature, we may approximate
their solution by setting $1 + M +s_1 \approx 1$ for 
$l < l_0 \sim \frac12 \ln (1/\sigma_1)$. Beyond that point, the
couplings simply evolve according to their simple scaling behavior at
the Gaussian fixed point. The equations may now be integrated, with
the result for the dimensionless renormalized DP mass 
\begin{equation}
 \label{renDPm}
	M(l_0) - s_1(l_0) = \sigma_1 \, e^{2l_0} \left( 1 - 2 \, 
	\exp \left[ - \int_0^{l_0} \ell(l') dl' \right] \right) \ .
\end{equation}
We may now distinguish various cases. For $d < 2$ and sufficiently
small $\sigma_1$ (large $l_0$), the exponent in Eq.~(\ref{renDPm})
behaves like $\ell^* l_0 \ll 1$, so that the DP mass is positive,
consistent with the system being in the inactive phase. For large
$\sigma_1$, however, the exponent is small and the DP mass is
negative. The critical point, in this approximation, occurs when
\begin{equation}
 \label{fp}
	\int_0^{l_0} \ell(l') \, dl'= \ln 2 \ .
\end{equation}
For small initial annihilation rate $\lambda$ we have
$\ell(l') \sim C_d \lambda e^{\epsilon l'}$. This gives $\sigma_{1c} 
\sim e^{-2 l_0} \sim (C_d \lambda / \epsilon \ln 2)^{2/\epsilon}$, in
agreement with Eq.~(\ref{sigmac}) apart from the factor of $\ln 2$,
which may be ascribed to the crudeness of the truncation procedure
described above. Thus, as expected, the one--loop RG equations
correctly sum the most singular diagrams in $1/\epsilon$ in the weak
coupling limit. However, from Eq.~(\ref{fp}) we may also extract the
behavior at strong coupling, because then 
$\ell(l') \sim \ell^* = \epsilon$. In that case we find 
\begin{equation}
	\sigma_{1c} \sim e^{-2 \ln2 / \epsilon} \ ,
\end{equation}
where, once again, the precise numerical factor is not to be taken too
seriously.

In exactly two dimensions, on the other hand, 
$\int_0^{l_0} \ell(l') \, dl' \sim \ln (1 + C_d \lambda \l_0)$,
c.f. Eq.~(\ref{2dflol}), so that $\sigma_{1c} \sim e^{-4 \pi/\lambda}$,
in precise agreement with the summed bubbles calculation,
Eq.~(\ref{crisig}). The only corrections to this formula are seen to
come from higher order loops, but, since $\ell$ is now marginally
irrelevant, it should asymptotically become exact as $\lambda \to
0$. Finally, we may use the criterion Eq.~(\ref{fp}) to explore the
case $d > 2$. In that case we may write
\begin{equation}
	\int_0^{l_0} \ell(l') \, dl' = \int_{\ell(0)}^{\ell(l_0)} 
	{\ell \, d \ell \over(d \ell / dl')} =
	\ln \left( {|\epsilon| + \ell(0) \over |\epsilon| + \ell(l_0)}
							\right) \ .
\end{equation}
Since now $\ell(l_0) \to 0$ as $\sigma_1 \to 0$, we see that there is
a possibility of finding the inactive phase for sufficiently small,
but nonzero $\sigma_1$ only if 
$\ln [(|\epsilon| + C_d \lambda) / |\epsilon|] > \ln 2$, i.e., 
$C_d \lambda > |\epsilon|$. Once again, of course, one must question
the validity of the loop expansion in this region. However, the result
does indicate that the inactive phase does not exist for nonzero
$\sigma_1$ at weak coupling.

\subsection{Larger odd values of m}

The advantage of the RG loop expansion approach outlined above for the 
case $m = 1$ is that it may readily be extended to larger values of
$m$, where the identification and resummation of the most singular
diagrams is somewhat more difficult. The only case that we shall treat
explicitly here is $m = 3$. In that case we have a branching rate
$\sigma_3$, corresponding to a dimensionless renormalized coupling
$s_3$, but, as before, couplings $\sigma_1$ and $\mu$ are
generated. (Note that $\sigma_2$ is also generated, but only at higher
orders, and in any case this is irrelevant near $d = 2$.) Once again,
the one--loop diagrams are simple, see Fig.~\ref{mreno}(a,b,d) and
differ from those for $m = 1$ only by combinatorial factors. The RG
flow equations are 
\begin{eqnarray}
	{d \ell(l) \over dl} &=& - \beta_\ell = \ell \left( \epsilon -
		{\ell \over (1+M+s_1+s_3)^{2-d/2}} \right) \ , \\ 
	{d s_3(l) \over dl} &=& 2 s_3 -
		{6 \ell s_3 \over (1+M+s_1+s_3)^{2-d/2}} \ , \\
	{d s_1(l) \over dl} &=& 2 s_1 -
		{\ell s_1 \over (1+M+s_1+s_3)^{2-d/2}} +
	        {6 \ell s_3 \over (1+M+s_1+s_3)^{2-d/2}} \ , \\
	{d M(l) \over dl} &=& 2 M + 
		{\ell s_1 \over (1+M+s_1+s_3)^{2-d/2}} \ .
\end{eqnarray}
Notice that the mass term is now $M+s_1+s_3$. As before, we see that
this combination obeys a simple equation, and that it grows
exponentially like $e^{2l}$. Thus, once again, we may make the same
approximation and ignore it in the denominators up to 
$l_0 \sim \frac12 \ln (1/\sigma_3)$. The remaining equations may now
be solved by quadrature. We give the results only for the case 
$d = 2$. If we shift the fields we now find that the DP mass is 
$\Delta \equiv \mu - \sigma_1 - 3 \sigma_3$. The dimensionless
renormalized version of this quantity at scale $l_0$ then turns out to
be 
\begin{equation}
	M(l_0) - s_1(l_0) - 3 s_3(l_0) = \sigma_3 \, e^{2l_0}
	\left[ 1 - {12/5 \over 1 + \lambda l_0 / 2 \pi} - 
		{8/5 \over (1+ \lambda l_0 / 2 \pi)^6} \right] \ .
\end{equation}
Once again, this is positive for large $l_0$ (small $\sigma_3$),
indicating the existence of an inactive phase, while it is clearly
negative for sufficiently small large $\sigma_3$. The zero must be
found numerically, and occurs at $\lambda l_0 / 2 \pi \approx 1.42$,
corresponding to 
\begin{equation}
	\sigma_{3c} \sim e^{-5.68 \pi / \lambda} \ .
\end{equation}
This is expected to be asymptotically exact (apart from a possible
prefactor) as $\lambda \to 0$. Note that, compared with the case 
$m = 1$, the transition occurs at a much smaller value of the
branching rate. This is because the generation of the process 
$A \to \emptyset$ now happens only at higher order in the annihilation
rate $\lambda$, which is asymptotically small. This pattern of the
decrease of $\sigma_{mc}$ with increasing $m$ continues, and is
consistent with what is observed in simulations \cite{taktre}. These
calculations for larger $m$ may also be extended to the cases 
$d \not= 2$, with similar results as for $m = 1$.


\section{BARW with three--particle annihilation}
 \label{3ABARW}

We finally briefly discuss the case of BARW with $k = 3$, i.e., 
three--particle annihilation processes instead of the two--particle
annihilation $A + A \to \emptyset$ which has concerned us in the bulk
of this paper. In physical dimensions $d \geq 1$, BARW with any
higher value of $k$ will either be satisfactorily described by the
mean--field rate equation (\ref{mfrate}), according to the fact that
the annihilation vertex $\lambda_k$ has upper critical dimension 
$d_c = 2 / (k-1)$, c.f. Eq.~(\ref{scdim3}), or simply generate the
processes with lower $k$, see below.

We may of course largely follow the arguments given in the preceding
sections. First, via a combination of branching and annihilation the
branching processes with $m$ offspring particles (rate $\sigma_m$) now
generate all the lower branching processes $\propto \sigma_{m-3}$,
$\sigma_{m-6}$, etc., see Fig.~\ref{k3bub}(a). Clearly, the reactions
with lowest offspring number will be the most relevant ones, and
therefore we shall essentially have to discuss only the cases $m = 1$,
$m = 2$, and $m = 3$. For the latter, and equally for any $m = 3 l$,
$l = 1,2,\ldots$, the branching and annihilation reactions conserve
the particle number locally modulo $3$, and this symmetry precludes a
fluctuation--induced generation of spontaneous particle decay
processes, which might lead to a phase transition of the DP
universality class. In analogy with the situation for even $m$ in the
case $k = 2$, there might potentially arise a new non--trivial
universality class below some novel critical dimension $d_c' < 1$; yet
in one dimension one merely expects logarithmic corrections to the
mean--field behavior.

This leaves us with the cases $m = 2$ and $m = 1$ (or, more generally,
$m = 2 \ {\rm mod} \ 3$ and $m = 1 \ {\rm mod} \ 3$). Here, the above
special symmetry is not present, and via the processes depicted in
Fig.~\ref{k3bub}(b) and (c), respectively, the single--particle decay
$A \to \emptyset$ is generated. For $m = 2$, this already happens to
lowest order in the branching and annihilation vertices, 
${\cal O}(\sigma_2 \lambda_3)$, see Fig.~\ref{k3bub}(b). In addition,
the processes $A \to A + A$ and $A + A \to \emptyset$ also become
generated, but require diagrams of ${\cal O}(\sigma_2^2 \lambda_3)$
and ${\cal O}(\sigma_2^2 \lambda_3^2)$, respectively. This leads to
all the terms required for Reggeon field theory to represent the
effective action for the ensuing dynamic phase transition, provided
$\sigma_{2c} > 0$ (compare Sec.~\ref{odBARW}). Yet as already the
lowest--order diagram in Fig.~\ref{k3bub}(b) is (logarithmically)
divergent in one dimension, following the arguments given for the
odd--offspring BARW with $k = 2$ we expect such a non--trivial
transition to occur in $d \leq 1$, with the exponents of the DP
universality class. In fact, as the BARW with $k = 2$ and $m = 1$
becomes generated, albeit to some high order in the original
couplings, the DP transition may actually persist for any $d \leq 2$. 
For the case $m = 1$, the lowest--order diagram, 
${\cal O}(\sigma_1^2 \lambda_3)$, diverges already for $d \leq 2$
dimensions, and as again the process $A + A \to \emptyset$ is
generated via diagrams of order ${\cal O}(\sigma_1^4 \lambda_3^2)$,
this maps onto BARW with $k = 2$ and odd $m$ and hence one has to
expect a non--trivial phase transition of the DP universality class for
$d \leq 2$ dimensions.


\section{Summary and conclusions}
 \label{sumcon}

We have studied branching and annihilating random walks defined by the
combined processes $2 A \to \emptyset$ ($k = 2$), $A \to (m+1) A$ with
both even and odd offspring number $m$ in low dimensions, where
fluctuation effects and particle anticorrelations are crucial. We have
employed a field--theoretic representation of the corresponding
diffusion--limited reaction problem, derived from the classical master
equation. Apart from the continuum limit, the derivation of the action
requires no further approximations (Sec.~\ref{masfth}).

While the mean--field rate equation predicts the existence of an
active phase only for any nonzero value $\sigma_m$ of the branching
rate (Sec.~\ref{introd}), in low dimensions fluctuations lead to the 
emergence of a non--trivial inactive phase as well, and a dynamic
phase transition at a critical value $\sigma_c$ (for fixed 
annihilation rate $\lambda$ and diffusion constant $D$). As a
consequence of the local parity conservation in the case of even $m$,
which is reflected in a discrete symmetry in the action, both the
inactive phase and the critical properties at the nonequilibrium phase
transition differ drastically for even and odd $m$, respectively, as
was already known from numerical simulations
\cite{grassb,taktre,evmjen,norsim,mondim,odmjen}. Using diagrammatic
summations and renormalization group methods, we believe we have
provided at least a qualitatively satisfying understanding of these
different universality classes \cite{cartau}.

For odd $m$ and $d \leq 2$, {\em including} the borderline dimension
$d_c = 2$, fluctuations generate the spontaneous decay 
$A \to \emptyset$, as well as branching processes with all odd
offspring numbers smaller than $m$. This single--particle decay
provides a very effective mechanism counteracting the particle
production through branching, and induces a non--trivial dynamic phase
transition characterized by the critical exponents of the directed 
percolation universality class, as has been established in simulations
\cite{celaut,taktre,mondim,odmjen}. In the emerging inactive phase,
the spontaneous decay processes dominate, and thus the long--time
decay is {\em exponential}, i.e. {\em faster} than in the pure
annihilation model without the branching processes. For dimensions 
$d > 2$, we find $\sigma_c = 0$, and the system is essentially
described by mean--field theory. Thus the values of the critical
exponents display a discontinuous jump at $d_c = 2$ (for the upper
critical dimension of directed percolation is $d_c = 4$). Precisely in
two dimensions, we still predict a non--trivial transition with DP
exponents, and first indications for this have recently been found
\cite{odmjen}. Indeed, even the tendency that $\sigma_c$ should {\em
decrease} with increasing $m$ is at least qualitatively explained by
our results (Sec.~\ref{odBARW}).

While the case of odd $m$ is thus well understood now, there remain a
number of open questions for even offspring numbers. Similarly to the
situation for odd $m$, the even $m$ universality class is determined
by the lowest possible value $m = 2$. Our analysis shows that near two
dimensions, the branching rate always remains a relevant quantity
(i.e., an {\em active} phase governed by exponential correlations),
and thus necessarily $\sigma_c = 0$. In Sec.~\ref{2dBARW}, we have
evaluated the critical exponents to second  order in 
$\epsilon = 2 - d$, and have discussed the logarithmic corrections
arising at the critical dimension $d_c = 2$. On the other hand, a
direct and {\em exact} computation in one dimension
(Sec.~\ref{1dBARW}) establishes that the branching rate is actually
{\em irrelevant} in the limit $\lambda \to \infty$, which implies the
existence of an {\em inactive} phase only in this special situation,
governed by the power laws of the pure annihilation model (see also
Ref.~\cite{taktre}). Therefore a {\em new critical dimension}  
$1 < d_c' < 2$ must exist, below which there appears a dynamic phase
transition to a {\em power--law} inactive phase, see Fig.~\ref{coupd}.

Within the framework of a truncated loop expansion to one--loop order
(Sec.~\ref{evBARW}), we may describe the above scenario as a function
of dimension $d$ at least qualitatively; in this approximation, we
find $d_c' = 4/3$. Although the structure of the phase diagram seems
to be correct, the ensuing one--loop critical exponents in this rather
uncontrolled approximation are well off the numerically found values
\cite{taktre,evmjen,norsim,drorac}. Yet unfortunately, we were unable
to improve on our approximations, e.g. by extending the calculations
to the two--loop level. Also, the emergence of dangerous irrelevant
variables precludes the firm establishment of scaling relations
\cite{norsim} relating exponents in the active phase, say, and those
{\em at} the critical point. Although so far there has been no
evidence for this, the possibility of violations of such scaling laws 
should be carefully and thoroughly reinvestigated in numerical
simulations. It would also certainly be highly desirable to possibly
design a different renormalization scheme to address the even $m$
universality class. The desirability of going beyond the one--loop
approximation is even clearer in the inactive phase, where the density 
is expected to decay as $A / (D_R t)^{d/2}$, for $d < 2$, and $D_R$ is
a renormalized diffusion constant. To one--loop order, no such
renormalization takes place. It is therefore not possible to establish
postulated scaling laws relating the manner in which $D_R$ is supposed
to behave near the critical point to the other exponents defined in
the active phase. The inactive phase is also interesting in that it is
expected to exhibit a kind of spontaneous symmetry breaking as a
result of the parity invariance of the dynamics. That is, if we start
with a state with a even number of particles (equivalent to imposing
periodic boundary conditions on the corresponding Ising model), the
asymptotic state is the trivial absorbing state (fully ordered in the
Ising language). But if we begin with an odd number of particles this
cannot happen: instead we expect the system to evolve to a state with
a {\em finite} average total number of particles. Once again, one may
postulate scaling laws relating the power law divergence of this
number as the critical point is approached to other exponents, and it
would be very useful to have a more systematic derivation of these,
which is so far lacking in our analysis. The study of the nature of
the inactive phase seems to be particularly interesting, as numerical
studies appear to indicate the existence of unexpected simplifications
\cite{raczpcomm}.

As discussed in Sec.~\ref{NsBARW}, an obvious generalization to 
$N > 2$ particle `flavors' leads to qualitatively new behavior,
governed by the {\em exactly} computable exponents of the limit 
$N \to \infty$. To our knowledge, this case of BARW with even $m$ and
$N > 2$ particle types has not been studied in simulations, nor have
BARW with $k = 3$ (see Sec.~\ref{3ABARW}).

\acknowledgments

We benefited from discussions with M.~Droz, G.~Grinstein, M.J.~Howard,
I.~Jensen, N.~Menyh\'ard, G.~\'Odor, K.~Oerding, Z.~R\'acz, and
G.M.~Sch\"utz.
This research was supported by the Engineering and Physical Sciences
Research Council (EPSRC) through Grant GR/J78327. 
U.C.T. acknowledges support from the European Commission through a TMR
Marie Curie Fellowship, contract No.~ERB FMBI-CT96-1189.

\begin{figure}
\caption{Elements of the perturbation theory for BARW with $k = 2$:
	(a) propagator $(-i \omega + D q^2 + \sigma_m)^{-1}$;
	(b,c) annihilation vertices $\lambda$ and $- \lambda$;
	(d) branching vertex $\sigma_m$; and 
	(e) pair production vertex $\tau$.}
 \label{mvert}
\end{figure}

\begin{figure}
\caption{One--loop diagrams for the renormalizations of
	(a) the annihilation vertex $\lambda$,
	(b) the branching vertex $\sigma_m$, and 
	(c) the pair production vertex $\tau$; 	
	(d) the generation of the branching process 
	    $\propto \sigma_{m-2}$ through a combination of branching 
	    $\propto \sigma_m$ and two--particle annihilation.}
 \label{mreno}
\end{figure}

\begin{figure}
\caption{UV--singular diagrams (near $d_c = 2$) to two--loop order for
	the vertex functions 
	(a) $\Gamma_{{\hat \psi} {\hat \psi} \psi \psi}$,
	(b) $\Gamma_{{\hat \psi} {\hat \psi} {\hat \psi} \psi}$, and
	(c) $\Gamma_{{\hat \psi} {\hat \psi}}$.}	
 \label{2dren}
\end{figure}

\begin{figure}
\caption{An important set of diagrams contributing to $P_{m+1}(t)$,
	i.e., to the renormalization of the branching process, for 
	$m = 2$.}
 \label{zgzag}
\end{figure}

\begin{figure}
\caption{Complete set of tree and one--loop diagrams for BARW with
	$m=2$, for the two-- and four--point vertex functions
	(a) $\Gamma_{\psi \psi}$,
	(b) $\Gamma_{{\hat \psi} \psi}$,
	(c) $\Gamma_{{\hat \psi} {\hat \psi}}$ (to first order in
            $\tau_0$), 
	(d) $\Gamma_{{\hat \psi} {\hat \psi} \psi \psi}$, and
	(e) $\Gamma_{{\hat \psi} {\hat \psi} {\hat \psi} \psi}$.}
 \label{1loop}
\end{figure}

\begin{figure}
\caption{Flow diagram $s^{2-d/2}$ vs. $\ell$ for BARW with $m=2$ in
	$d = 1$. For the flows depicted here, the initial values for
	the couplings are $\ell(0) = 0.01$, with (clockwise)
	$s(0) = 0.01$ (long--dashed), $s(0) = 0.0005$ (dotted),
	$s(0) = 0.00025$ (dashed), $s(0) = 0.0002$ (solid),
	$s(0) = 0.00015$ (dotted), $s(0) = 0.0001$ (dashed),
	and $s(0) = 0.00005$ (solid).}
 \label{fld=1}
\end{figure}

\begin{figure}
\caption{Phase diagram for BARW with even $m$ as function of 
	dimension. \qquad\qquad\qquad\quad$\;$}
 \label{coupd}
\end{figure}

\begin{figure}
\caption{Bethe--Salpeter equations for
	(a) the annihilation rate $\lambda$,
	(b) the branching rate $\sigma'$, and
	(c) the pair production rate $\tau$ 
	of the $N$ species model in the limit $N \to \infty$.}
 \label{nspec}
\end{figure}

\begin{figure}
\caption{Most singular (bubble) diagrams for the renormalization of
	(a) the particle decay rate $\mu$,  
	(b) the branching rate $\sigma_1$, and
	(c) the propagator (mass) $\mu + \sigma$
	for BARW with $m = 1$ near $d_c = 2$ dimensions.}
 \label{buble}
\end{figure}

\begin{figure}
\caption{Diagrams for BARW with $3 A \to \emptyset$ for the generation
	(a) of the branching process $\propto \sigma_{m-3}$ from
	    $\sigma_m$ and $\lambda_3$, 
	(b) of the particle decay $A \to \emptyset$ for the case
	    $m = 2$, and 
	(c) for $m = 1$.}
 \label{k3bub}
\end{figure}

\newpage

{\bf FIG.~\ref{mvert} \qquad (J.L.~Cardy and U.C.~T\"auber) \\ \\ \\}
\begin{figure}
\epsfxsize = 5.5 truein
\epsffile{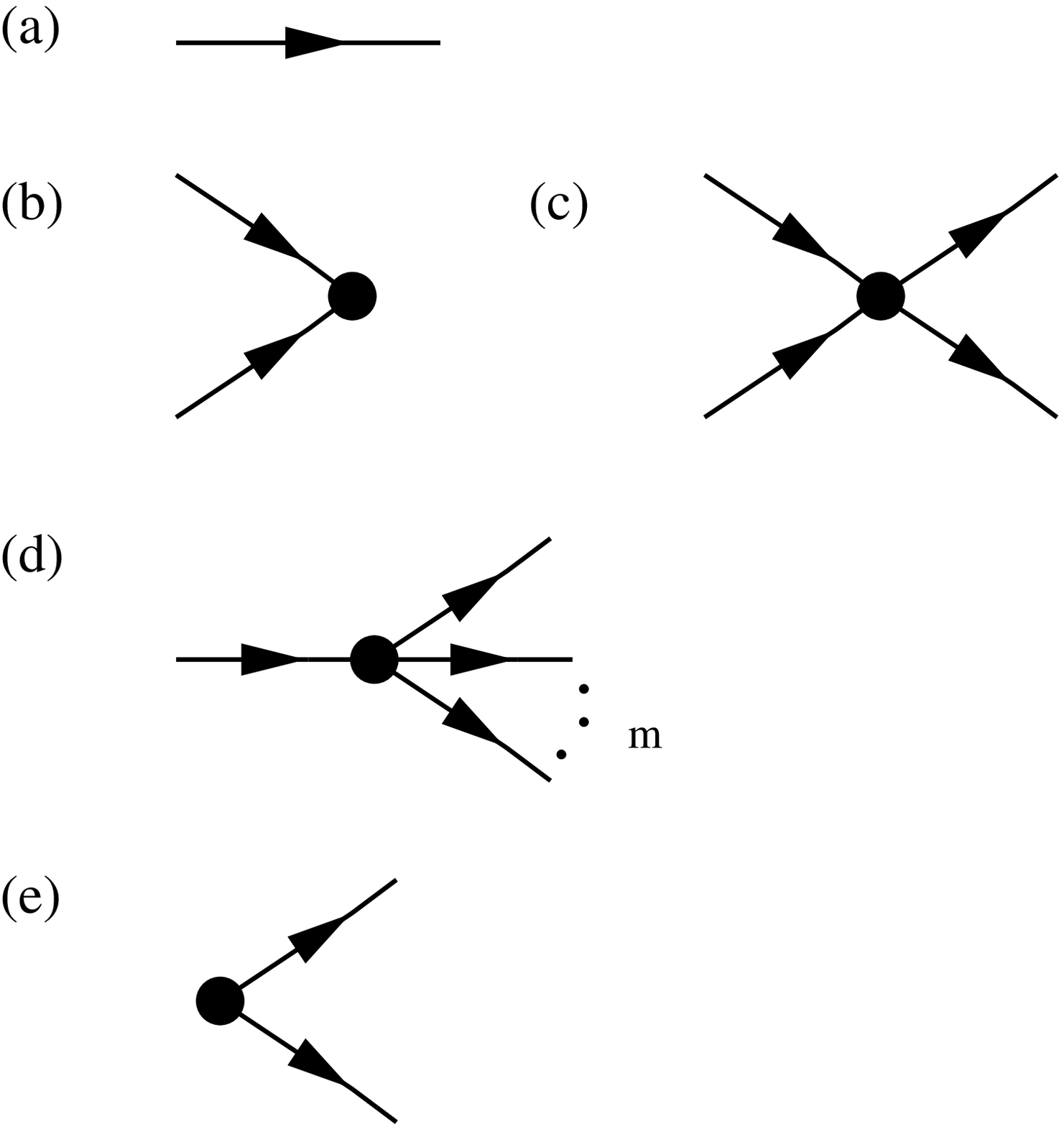}
\end{figure}
\newpage

{\bf FIG.~\ref{mreno} \qquad (J.L.~Cardy and U.C.~T\"auber) \\ \\ \\}
\begin{figure}
\epsfxsize = 6.0 truein
\epsffile{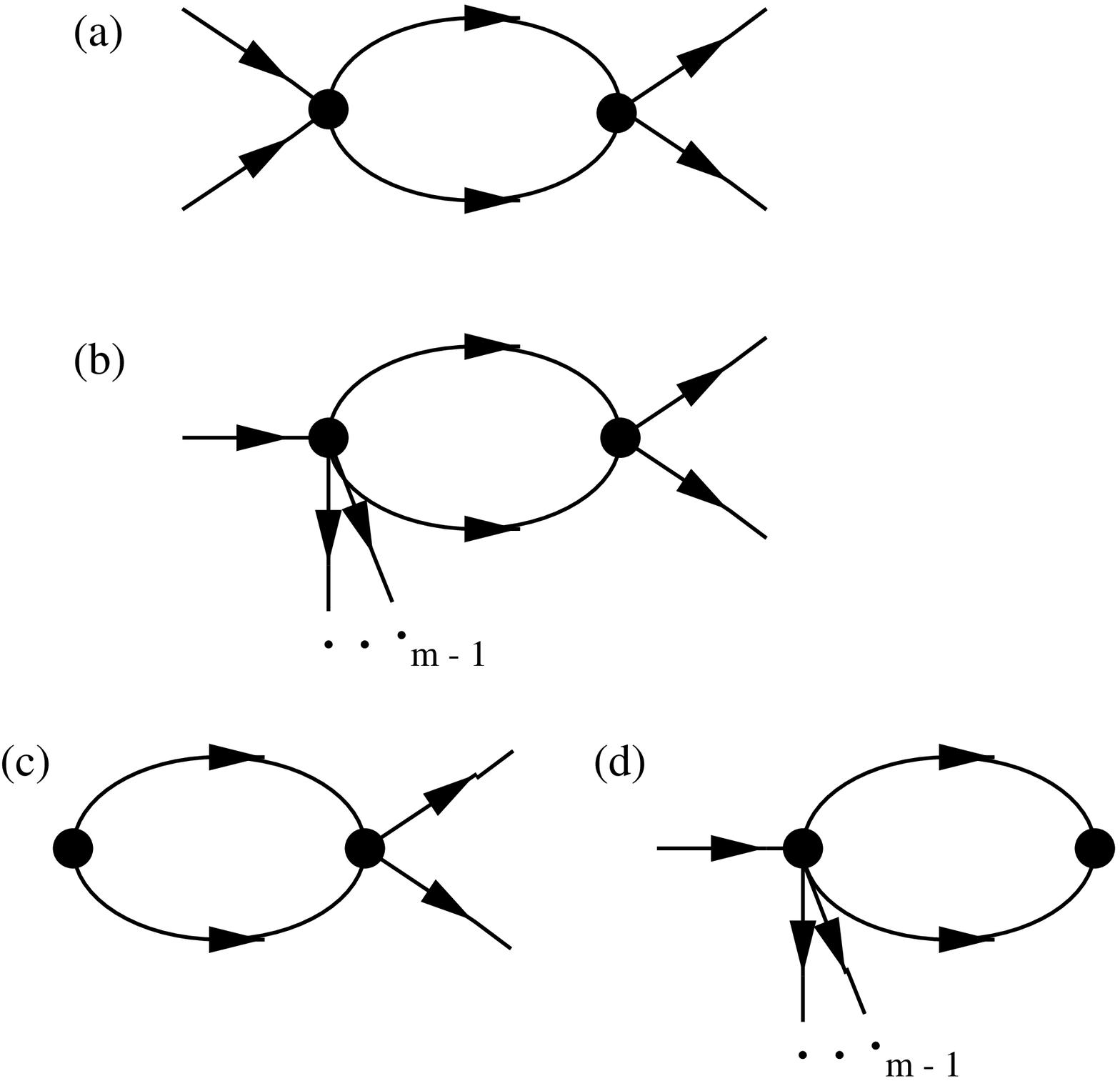}
\end{figure}
\newpage

{\bf FIG.~\ref{2dren} \qquad (J.L.~Cardy and U.C.~T\"auber) \\ \\} 
\begin{figure}
\epsfxsize = 5 truein
\epsffile{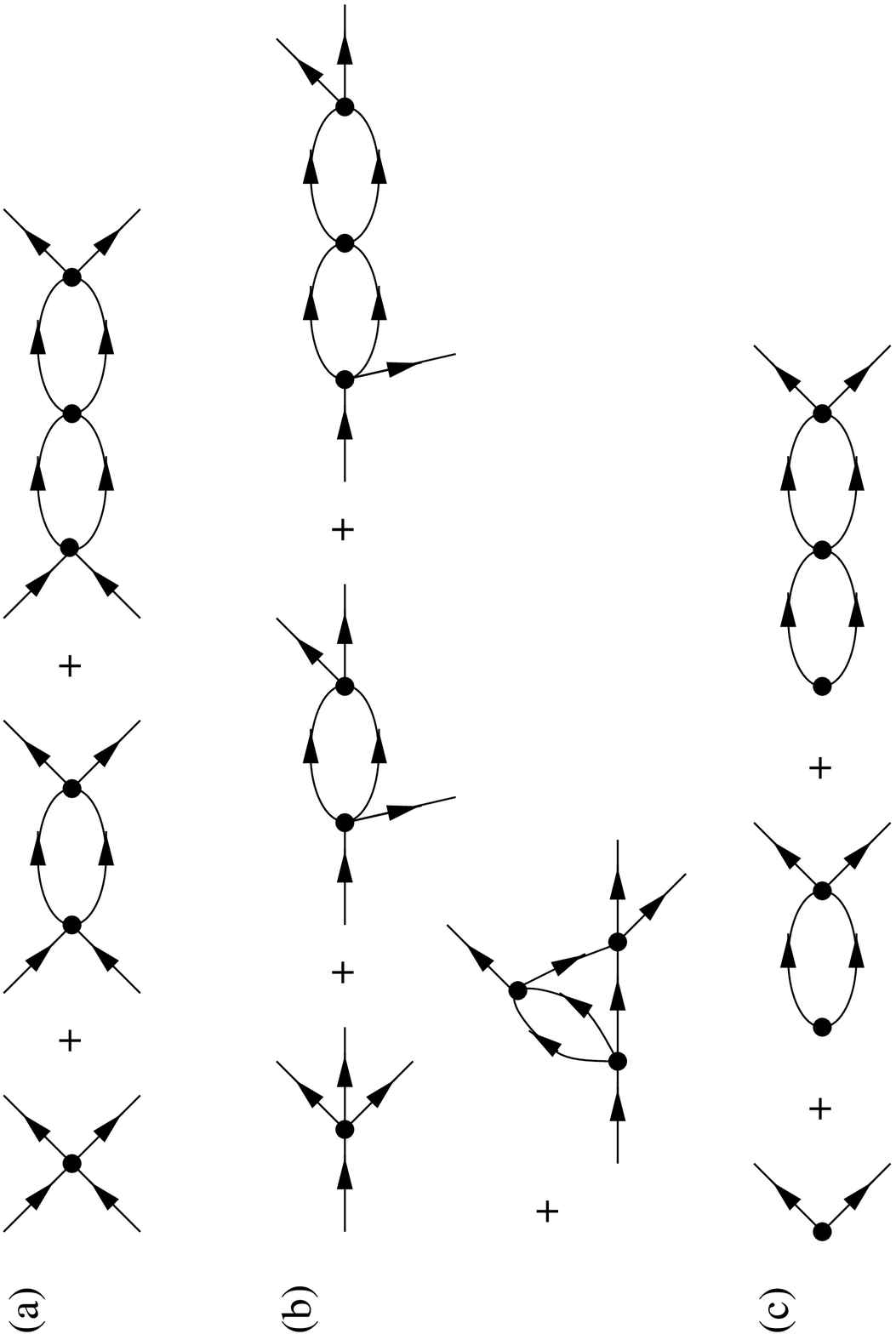}
\end{figure}
\newpage

{\bf FIG.~\ref{zgzag} \qquad (J.L.~Cardy and U.C.~T\"auber) \\ \\ \\} 
\begin{figure}
\epsfxsize = 5 truein
\epsffile{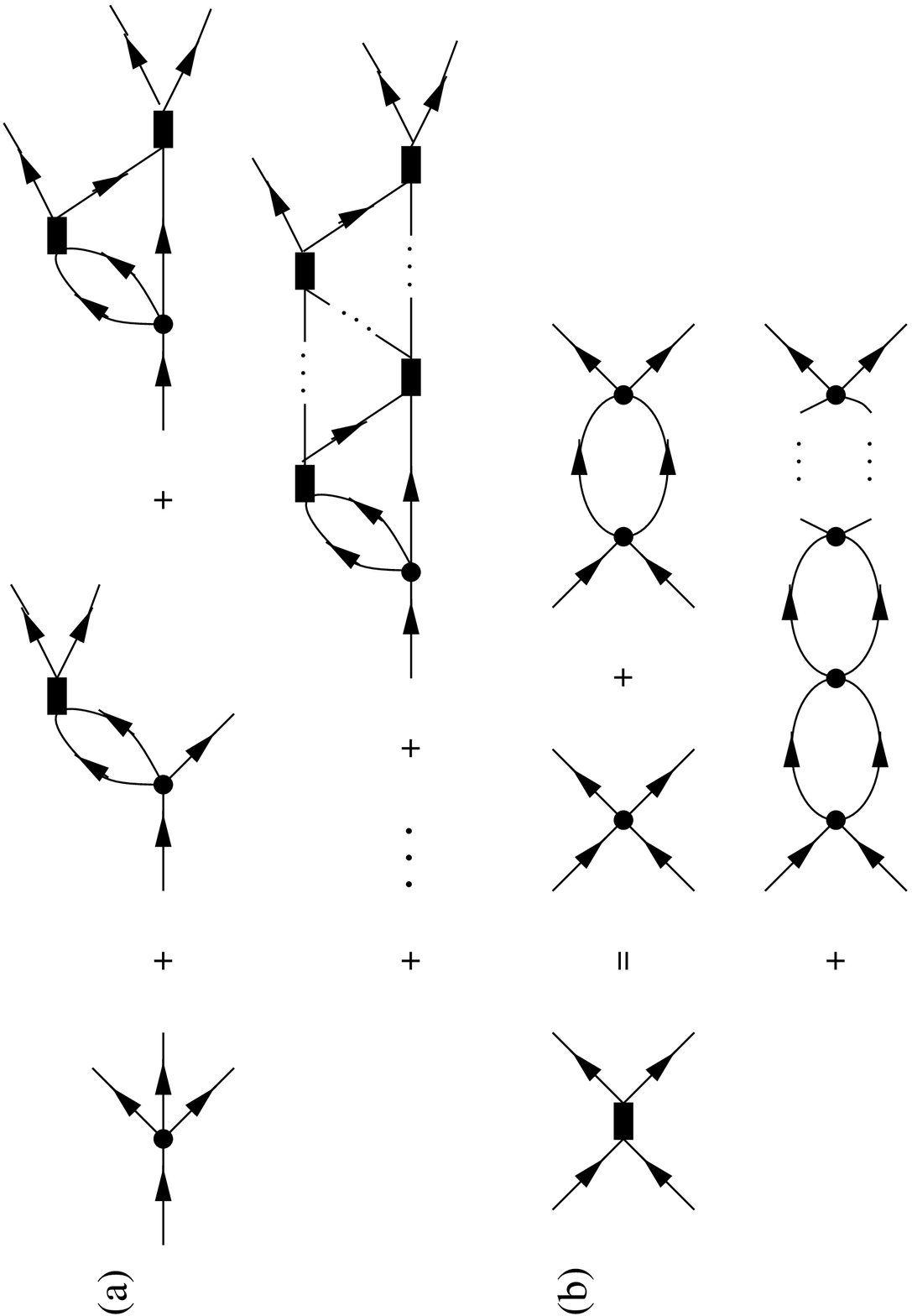}
\end{figure}
\newpage

{\bf FIG.~\ref{1loop} \qquad (J.L.~Cardy and U.C.~T\"auber) \\ \\ \\} 
\begin{figure}
\epsfxsize = 6.0 truein
\epsffile{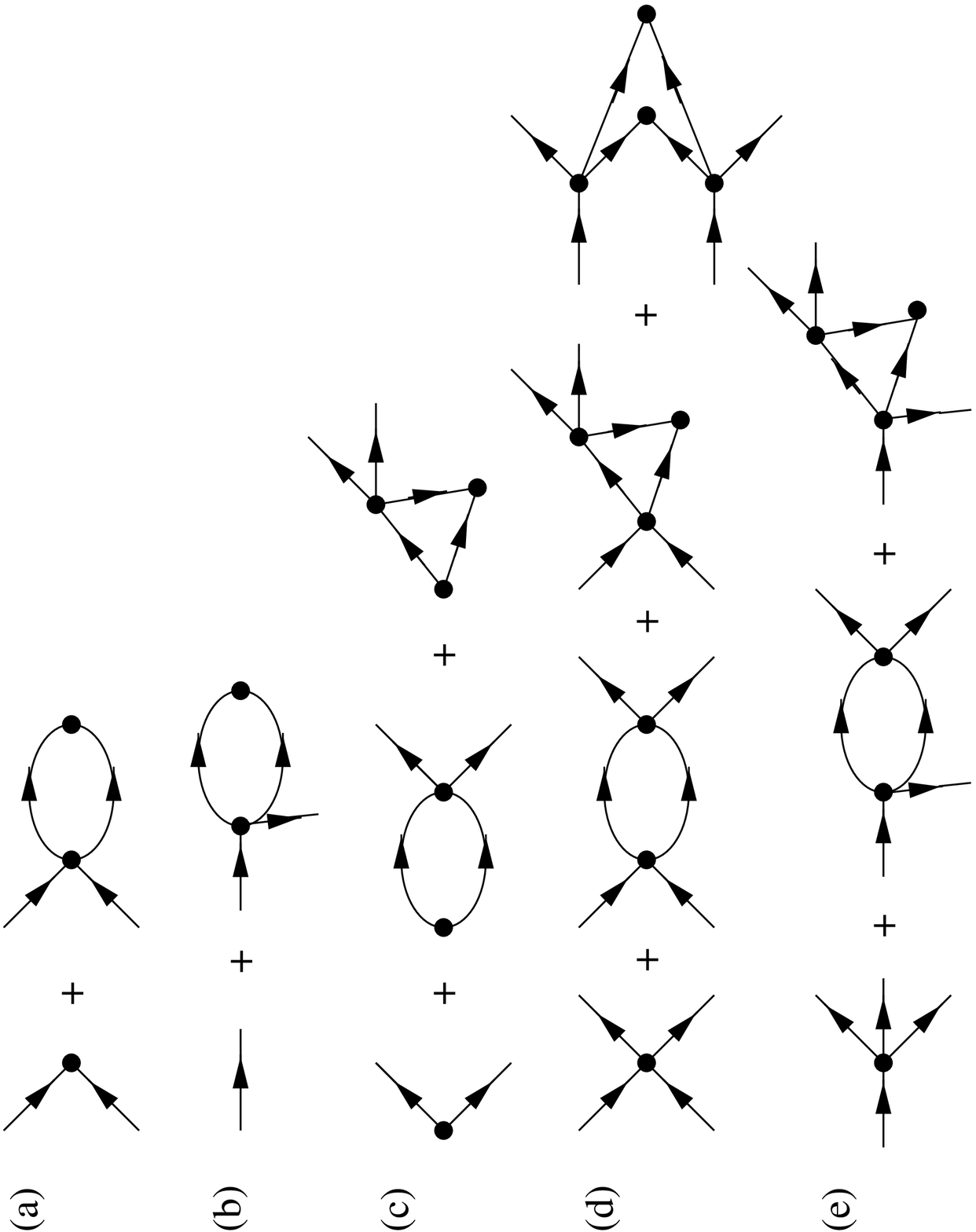}
\end{figure}
\newpage

{\bf FIG.~\ref{fld=1} \qquad (J.L.~Cardy and U.C.~T\"auber) \\ \\ \\}
\begin{figure}
\epsfxsize = 6.0 truein
\epsffile{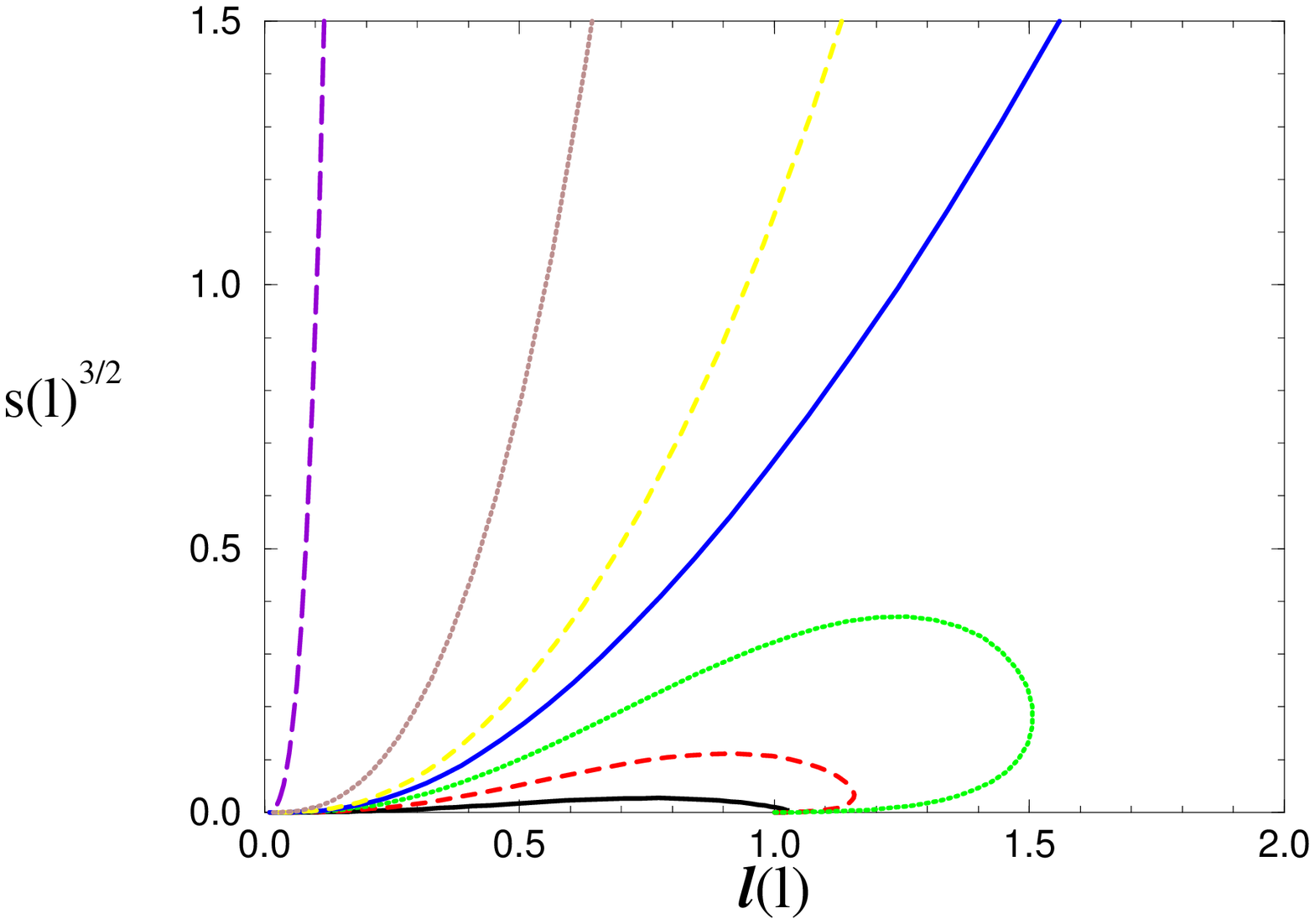}
\end{figure}
\newpage

{\bf FIG.~\ref{coupd} \qquad (J.L.~Cardy and U.C.~T\"auber) \\ \\ \\} 
\begin{figure}
\epsfxsize = 5.0 truein
\epsffile{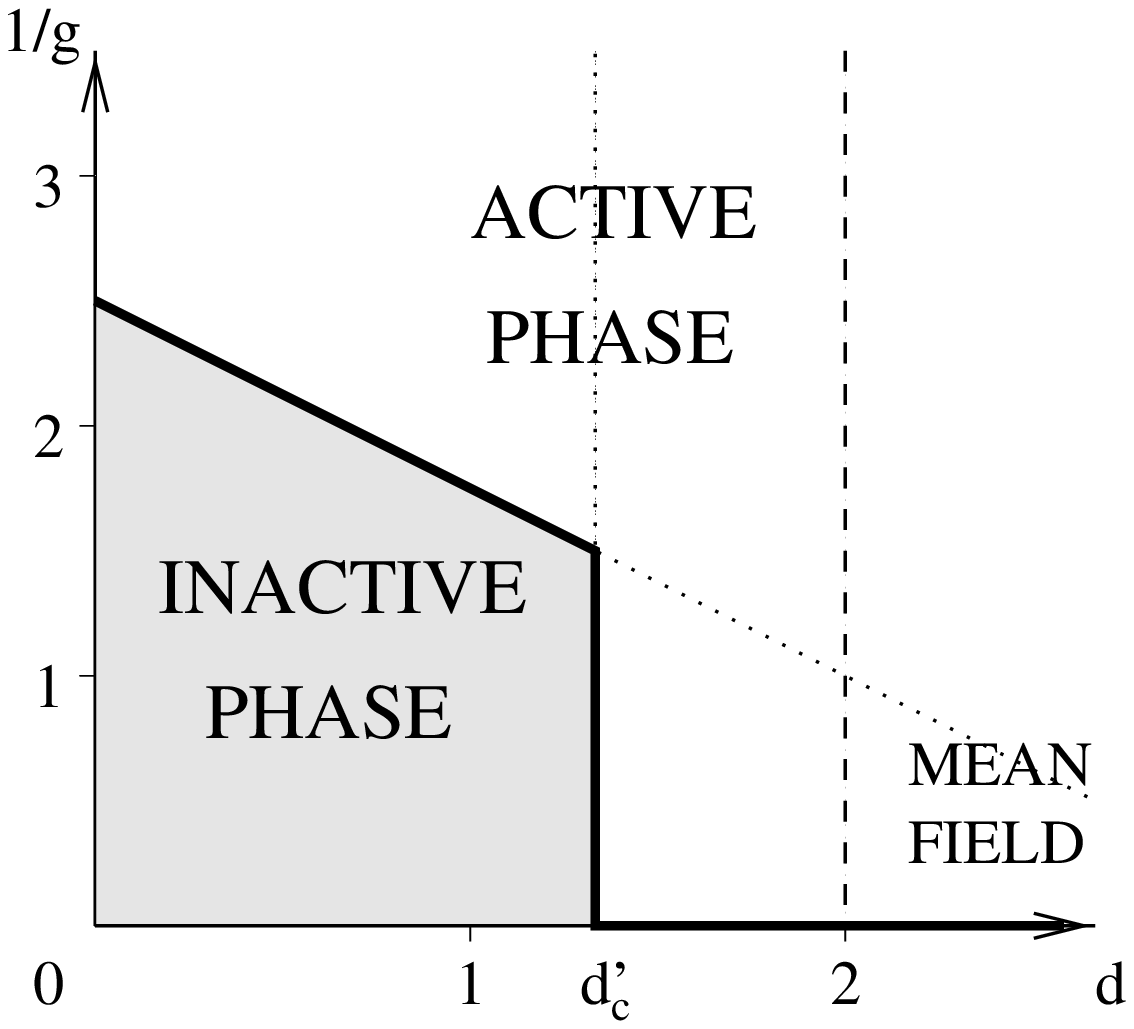}
\end{figure}
\newpage

{\bf FIG.~\ref{nspec} \qquad (J.L.~Cardy and U.C.~T\"auber) \\ \\ \\} 
\begin{figure}
\epsfxsize = 5.0 truein
\epsffile{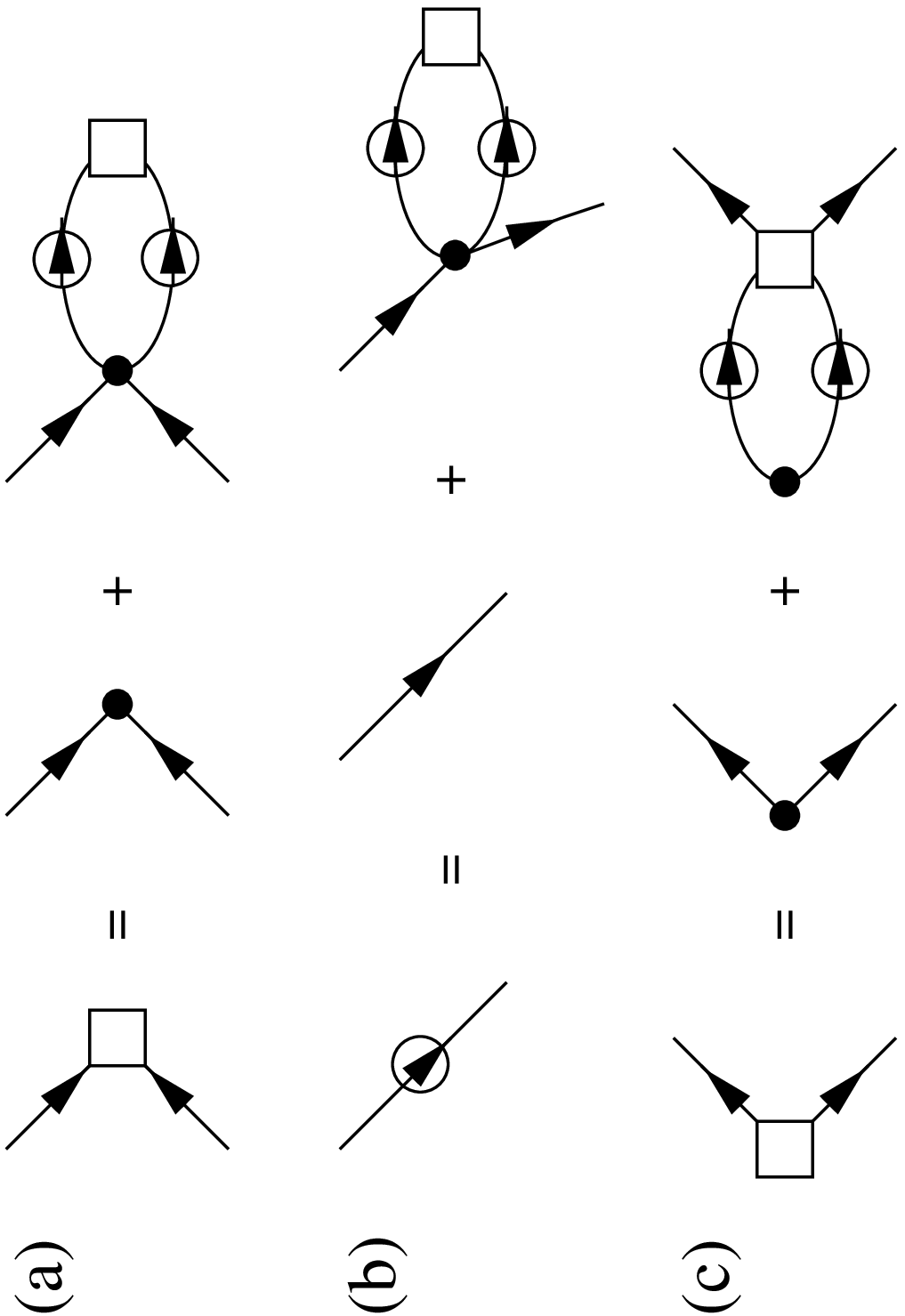}
\end{figure}
\newpage

{\bf FIG.~\ref{buble} \qquad  (J.L.~Cardy and U.C.~T\"auber) \\ \\ \\} 
\begin{figure}
\epsfxsize = 5.5 truein
\epsffile{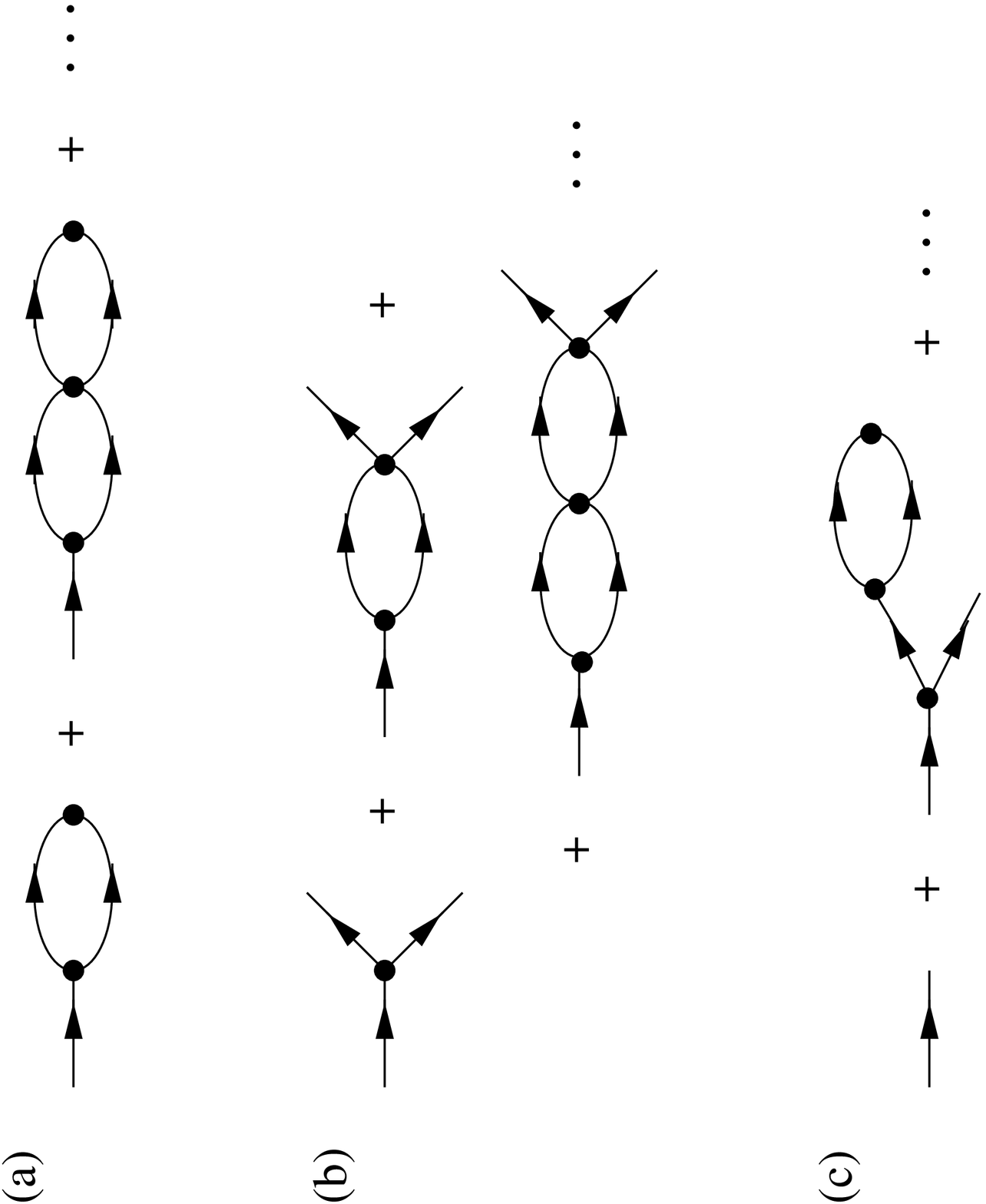}
\end{figure}
\newpage

{\bf FIG.~\ref{k3bub} \qquad (J.L.~Cardy and U.C.~T\"auber) \\} 
\begin{figure}
\epsfxsize = 4.5 truein
\epsffile{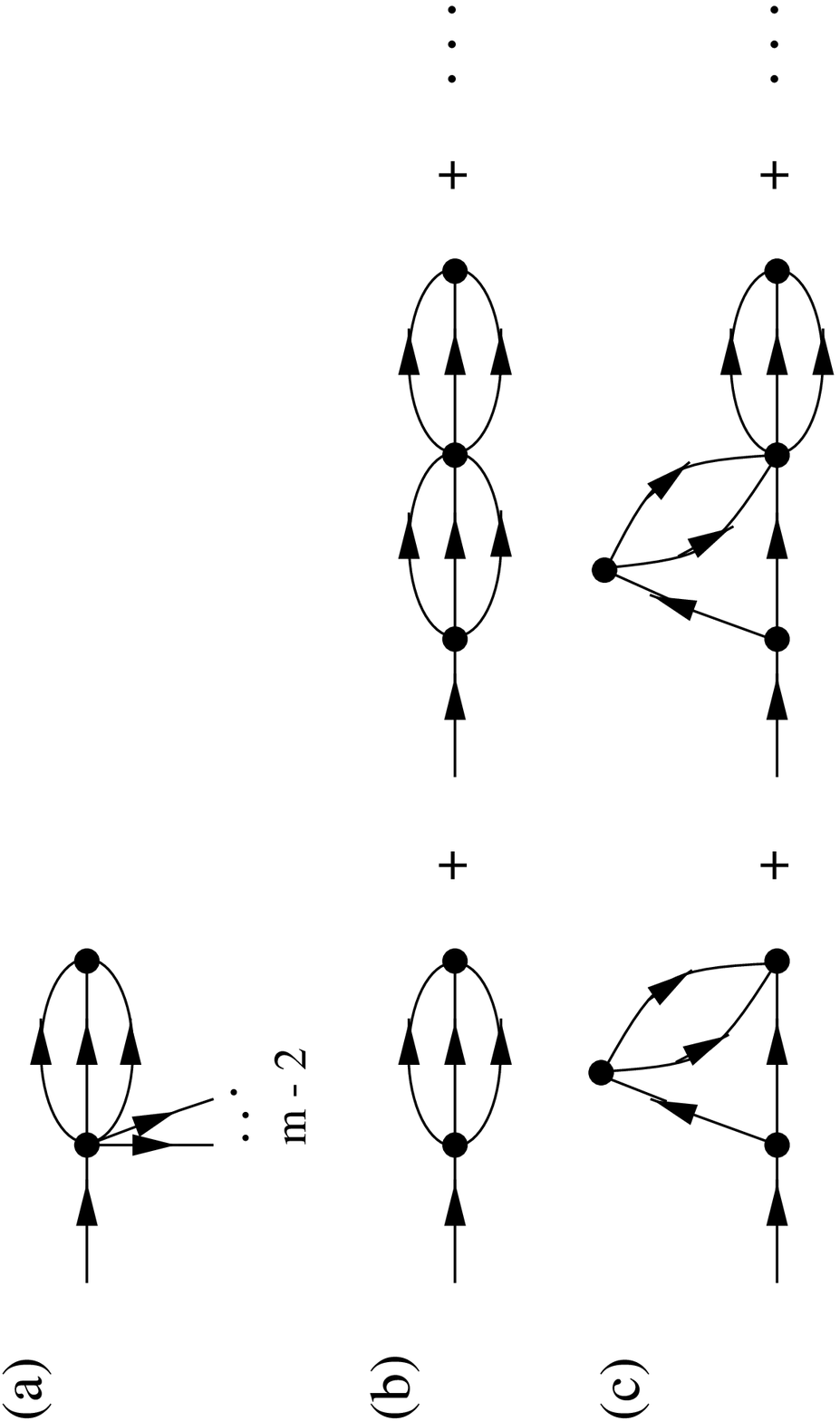}
\end{figure}


\end{document}